\documentclass[
aps,
preprint,
showpacs,
nofootinbib,
groupedaddress]{revtex4}

\usepackage{graphicx} 
\usepackage{amsmath} 
\usepackage{amssymb} 
\usepackage{pstricks,pst-node,pst-coil} 
\def\identity{{\rm 1\kern -.23em l}} 
\def\det{{\mathrm{det}}} 
 
\def\tr{{\mathrm{tr}}} 

\begin{document}

\title{One-loop fermionic corrections to the instanton transition in two dimensional chiral Higgs model} 
\author{Y. Burnier}
\email{Yannis.Burnier@epfl.ch}
\affiliation{
Institut de Th\'eorie des Ph\'enom\`enes Physiques,
Ecole Polytechnique F\'ed\'erale de Lausanne,
CH-1015 Lausanne, Switzerland}
\author{M. Shaposhnikov}
\email{Mikhail.Shaposhnikov@epfl.ch}
\affiliation{
Institut de Th\'eorie des Ph\'enom\`enes Physiques,
Ecole Polytechnique F\'ed\'erale de Lausanne,
CH-1015 Lausanne, Switzerland}

\date{July 11, 2005}

\begin{abstract}
The one-loop fermionic contribution to the probability of an 
instanton transition with fermion number violation is calculated in 
the chiral Abelian Higgs model in 1+1 dimensions, where the fermions 
have a Yukawa coupling to the scalar field. The dependence of 
the determinant on fermionic, scalar and vector mass is determined. We 
show in detail how to renormalize the fermionic determinant  in 
partial wave analysis, which is convenient for computations. 
\end{abstract}
\pacs{11.15.Kc, 11.30.Fs, 11.30.Rd}
\maketitle 

\section{Introduction} 
 
The interest in the chiral Abelian Higgs model in 1+1 dimensions lies in 
the fact that it shares some properties with the electroweak 
theory, but is much simpler and may serve as a toy model. One of the 
most interesting common features of the two gauge theories is the fermionic number  
non-conservation \cite{thooft}. Both give rise to instanton 
transitions, leading to the creation of a net fermion number due to an anomaly 
\cite{Adler:1969gk,Bell:1969ts}. Both theories contain finite 
temperature sphaleron transitions \cite{Kuzmin:1985mm} - 
\cite{Bochkarev:1989vu}.   
 
At zero temperature, zero fermionic chemical potentials and for a  
small number of particles participating in the reaction, the 
probability of the process can be computed using semi-classical 
methods. In general,
 the result is a product of the exponential of 
the classical action  $e^{-S_{cl}}$ and the fluctuation determinants. 
The latter factor includes the small perturbations of the fields around 
the instanton configuration, and in many cases may only be computed 
numerically. 
 
Quite a number of computations of determinants in 1+1 dimensions can 
be found in the literature\footnote{For 3+1 dimensional computation without 
Yukawa couplings see the seminal paper by 't Hooft \cite{thooft}.}.  
In particular, the determinants have been calculated for the vector 
and scalar field fluctuations around the instanton in \cite{detAphi}, 
as well as for the fermionic ones in \cite{detpsim0}, where it was 
assumed that fermions have no mass term and no interaction with the 
Higgs field. However, to our best knowledge, no computations 
incorporating the Yukawa coupling of the fermions to scalar field  have been 
done till now, neither for realistic case of electroweak theory nor 
for the chiral Abelian Higgs model \footnote{The determinants in the 
high temperature sphaleron transition in 1+1 dimensions were computed in 
\cite{Bochkarev:1987wg,Bochkarev:1989vu}.}.  
 
The aim of the present work is to partially fill this gap, 
calculating the fermionic determinant in the 1+1 dimensional case, 
where the fermions interact with the Higgs field in a similar way as 
in the electroweak theory \footnote{Similar studies have 
recently been performed for other models. In a supersymmetric theory 
in 2+1 dimensions the calculation is simplified by a supersymmetric 
constraint \cite{susyvortex}. The fermionic contribution to the 
vortex mass has been calculated in a model resembling the (non 
chiral) Abelian Higgs gauge theory, where the fermion couples to the 
absolute value of the scalar field \cite{|phi|}.}.
 
This calculation is somewhat delicate because of the difficulties 
occurring in regularization and renormalization of chiral gauge 
models beyond perturbation theory. Furthermore, an analytic solution 
to this problem cannot be obtained, since even the classical 
instanton profile, given by the Nielsen-Olesen string solution 
\cite{corde}, is not known analytically, apart from the special case 
where the Higgs mass equals the vector field mass \cite{cordean}. 
Nevertheless, we use analytical methods as long as possible 
 before moving on
to numerical computation. We will use a numerical method 
developed in \cite{BK}, extended to our case. 
 
The paper is organized as follows. In section 2, the model and its 
basic features, such as its vacuum structure, anomaly, instanton 
configuration and fermionic zero modes are discussed. In section 3, 
we study  and compare the 1-loop divergences occurring in this model in 
various regularization schemes. In section 4, the method of \cite{BK} 
to calculate determinants is discussed and applied to our case. 
In section 5 we present in some detail 
the numerical procedures and give the results of the determinant 
computation. Finally, conclusions are given in section 6. 
 
\section{The model} 
 
The model we consider here contains a complex scalar field $\phi$ 
with vacuum expectation value $v$; a vector field $A_\mu$, and  
$n_f$ fermions $\Psi^j$, $j=1,...,n_f$: 
\begin{eqnarray} 
\mathcal{L} &=&-\frac{1}{4}F^{\mu \nu }F_{\mu \nu }+i\overline{\Psi 
}^j\gamma ^{\mu }(\partial _{\mu }-i\frac{e}{2}\gamma ^{5}A_{\mu 
})\Psi^j  \notag \\ &&-V(\phi )+\frac{1}{2}\left| D_{\mu }\phi 
\right| ^{2}+if^j\overline{\Psi}^j\frac{1+\gamma_5}{2}\Psi^j\phi^*  
-if^j\overline{\Psi}^j\frac{1-\gamma_5}{2}\Psi^j\phi.  
\label{Lm} 
\end{eqnarray} 
The charges of the left and right-handed fermions differ by a sign, 
$e_L=-e_R=\frac{e}{2}$ and the symmetry breaking potential is chosen 
to be $V[\phi]=\frac{\lambda}{4}(|\phi|^2-v^2)^2$. In the following, we
 use the Majorana
representation for the $\gamma$-matrices: 
\begin{equation}
\gamma _{0}=\left( 
\begin{array}{cc}
0 & -i \\ 
i & 0
\end{array}
\right) ,\text{ }\gamma _{1}=\left( 
\begin{array}{cc}
0 & i \\ 
i & 0
\end{array}\right),~\gamma_5=\gamma_0\gamma_1.
\end{equation}
Note that we do not reduce generality in 
considering Yukawa interaction between identical fermions only\footnote{A 
more general interaction could  be written in the form 
$i\tilde{f}^{ij}\overline{\Psi}^i\frac{1+\gamma_5}{2} \Psi^j\phi^* 
+h.c.$ but the matrix $\tilde{f}^{ij}$ can always be diagonalized 
and made real trough redefinition of the fields $\Psi^j$.}. In principle 
another mass term of the form $M\Psi^T\gamma_0\Psi+h.\,c.$ 
could be added to the Lagangian (\ref{Lm}). It is compatible with gauge and 
Lorentz invariance but breaks fermion
number explicitly. As we are interested in instanton mediated 
fermion number non-conservation, we will not consider this term. 

This model has been studied as a toy model for the fermionic number 
non-conservation in the electroweak theory in a number of papers, see, 
e.g. \cite{Grigoriev:1988bd}-\cite{detAphi}, \cite{Ringwald}. 
 
The particle spectrum consists of a Higgs field with mass 
$m_H^2=2\lambda v^2$, a vector boson of mass $m_W=ev$, and $n_f$ Dirac 
fermions acquiring a mass $F^j=f^jv$ via Yukawa coupling. The model is 
free from gauge anomaly. There is, however, a chiral anomaly leading to the
non-conservation of the fermionic current, 
\[ 
J_\mu=J_\mu^L+J_\mu^R=\sum_{j=1}^{n_f} 
\overline{\Psi}^{j}_L\gamma_\mu\Psi^j_L+\sum_{j=1}^{n_f} 
\overline{\Psi}^{j}_R\gamma_\mu\Psi^j_R 
=\sum_{j=1}^{n_f} \overline{\Psi}^{j}\gamma_\mu\Psi^j, 
\]  
with a divergence given by 
\begin{equation} 
\partial_\mu 
J_\mu=\partial_\mu J_\mu^L+\partial_\mu J_\mu^R=-n_f\frac{e_L}{4\pi} 
\varepsilon_{\mu\nu}F_{\mu\nu}+n_f\frac{e_R}{4\pi}\varepsilon_{\mu\nu}F_{\mu\nu} 
= -n_f\frac{e}{4\pi}\varepsilon_{\mu\nu}F_{\mu\nu}. 
\label{curentanom} 
\end{equation} 
 
The vacuum structure of this model is non-trivial 
\cite{Jackiw:1976pf}. Taking the $A_0=0$ gauge and putting the theory 
in a spatial box of length $L$ with periodic boundary conditions, one 
finds that there is an infinity of degenerate vacuum states  
$|n\rangle,~ n \in \mathbf{Z}$ with the gauge-Higgs configurations 
given by 
\begin{equation} 
A_1= \frac{2\pi n}{eL},\quad \phi=ve^{i\frac{2\pi nx}{L}}. 
\label{vacst} 
\end{equation} 
 
The transition between two neighboring vacua, described by an instanton,   
leads to the non-conservation of fermion number by $n_f$ units. In this 
paper we consider $n_f$ to be even. The case of odd $n_f$, resulting in the
creation of an {\em odd} number of fermions, is analyzed in 
\cite{future}. 
 
\subsection{Lagrangian in Euclidean space} 
 
As the tunneling is best described in Euclidean space-time, we review 
here the corresponding equations and conventions.  
 
The Lagrangian (\ref{Lm}) may be rewritten in Euclidean space: 
\begin{eqnarray} 
\mathcal{L}^{E} &=&\frac{1}{4}F_{\mu \nu }F_{\mu \nu 
}+i\overline{\Psi}^{j} \gamma _{\mu }^{E}(\partial _{\mu 
}-i\frac{e}{2}\gamma _{5}A_{\mu })\Psi^j +V(\phi )\notag \\ 
&&+\frac{1}{2}(D_{\mu }\phi )^{\dagger }(D_{\mu }\phi 
)-if^j\overline{\Psi}^j\frac{1+\gamma_5}{2}\Psi^j\phi^*  
+if^j\overline{\Psi}^j\frac{1-\gamma_5}{2}\Psi^j\phi,  
\label{L} 
\end{eqnarray} 
with $D_{\mu}=\partial _{\mu}-ieA_{\mu}$, $\gamma_0^E=i\gamma_0$ and $\gamma_1^E=\gamma_1$. The fields $\overline{\Psi 
}$ and $\Psi$ are independent variables, and the gauge transformation 
reads: 
\begin{eqnarray} 
\Psi  &\longrightarrow &e^{i\alpha (x)\frac{\gamma _{5}}{2}}\Psi 
\text{ , } 
\overline{\Psi}  \longrightarrow 
\overline{\Psi }e^{i\alpha (x)\frac{\gamma _{5}}{2}},\notag \\ 
\phi  &\longrightarrow &e^{i\alpha (x)}\phi. 
\end{eqnarray} 
For comparison, the Lorentz transformation is: 
\begin{eqnarray*} 
\Psi(x)  &\rightarrow & \Psi'(x')= \Lambda_s \Psi(\Lambda^{-1} x'),  
\\ \overline{\Psi }(x) &\rightarrow 
&\overline{\Psi}'(x')=\overline{\Psi }\Lambda_s ^{-1}(\Lambda^{-1} 
x'),  
\end{eqnarray*} 
with $\Lambda_s=\exp(i\gamma^5\frac{\theta}{2})$ being the rotation 
matrix in two dimensions.   
 
\subsection{Instanton} 
 
The instanton which describes the tunneling between the states 
$|0\rangle$ and $|n\rangle$ is simply the Nielsen-Olesen vortex with 
winding number $n$ \cite{corde}, which is a solution of the Euclidean 
equations of motion in two dimensions. In 
polar coordinates $(r,~\theta )$, the field configuration reads:  
\begin{eqnarray} 
\phi (r,\theta ) &=&e^{in\theta }\phi(r)=e^{in\theta }vf(r)  
\label{phi}, \\ A^{i}(r,\theta ) &=&\varepsilon 
^{ij}\widehat{r}^{j}A(r)   
\label{A}, 
\end{eqnarray}
where $\widehat{r}$ is the unit 
vector $\widehat{r}=(\cos\theta,\ \sin\theta)$ and $\varepsilon 
^{ij}$ the completely antisymmetric tensor with $\varepsilon 
^{01}=1$.
The functions $A$ and $f$ have to satisfy the 
following limits:   
\begin{eqnarray} 
&&f(r)\overset{r\rightarrow 0}{\longrightarrow }c r^{\left| n\right| } 
\label{l1},\notag\\ 
&&f(r)\overset{r\rightarrow \infty }{\longrightarrow }1, 
\label{l2}\notag \\ 
&& A(r)\overset{r\rightarrow 0}{\longrightarrow }0, 
\label{l3}\\ 
&&A(r)\overset{r\rightarrow \infty }{\longrightarrow }-\frac{n}{er}. 
\label{l4} \notag
\end{eqnarray} 
Passing to dimensionless variables 
\begin{equation} 
A=\frac{m}{e}\tilde{A}, \quad \phi=\frac{m}{e}\tilde{\phi}, \quad 
r=\frac{\tilde{r}}{m}\quad \text{with}\quad m=\sqrt{\lambda v^2} \label{dimless}
\end{equation} 
reduces the number of free parameters. The equations for $\tilde{A}, \tilde{\phi}$ are : 
\begin{eqnarray}  
-\partial_{\tilde{r}}\left(\frac{1}{\tilde{r}}\partial_{\tilde{r}} \tilde{r} 
\tilde{A}(r)\right)+\tilde{\phi}^2\left(\tilde{A}(r)-\frac{1}{\tilde{r}}\right) 
&=& 0,\notag\\ -\frac{1}{\tilde{r}}\partial_{\tilde{r}}\left(\tilde{r}
\partial_{\tilde{r}}
\tilde{\phi}(r)\right)+\left(\left(\frac{1}{\tilde{r}}-\tilde{A}(r)\right)^2
-1+\mu^2 
\tilde{\phi}(r)^2\right)\tilde{\phi}(r)&=&0,\label{vortex2} 
\end{eqnarray} 
with $\mu=\frac{m_H^2}{2m_W^2}=\frac{\lambda}{e^2}$.
The classical action is given by:
\begin{eqnarray} 
S_{cl}&=&\pi v^2 \int_0^\infty 
\tilde{r}d\tilde{r}\left\{\mu^2\left(\tilde{A}'(r)+\frac{\tilde{A}(r)}{\tilde{r}}
\right)^2\right. 
\label{actionclassiquesansdim}\\ &+&\left.\mu^2\left(\tilde{\phi}'(r)^2+ 
\tilde{\phi}(r)\left(\tilde{A}(r)-\frac{1}{\tilde{r}}\right)^2\right)
+\frac{\mu^4}{2} 
\left(\tilde{\phi}^2(r)-\frac{1}{\mu^{2}}\right)^2\right\}.\notag 
\end{eqnarray} 
The number $\Delta N$ of fermions created in the instanton transition 
can be computed by integrating (\ref{curentanom}) over the Euclidean 
space: 
\begin{equation} 
\Delta N=-\int d^2x\partial_\mu 
J_\mu=-n_f\int d^2x \frac{e}{4\pi}\varepsilon_{\mu\nu}F_{\mu\nu}  
=-q n_f,\label{topcharge} 
\end{equation} 
where $q=\int d^2x \frac{e}{4\pi}\varepsilon_{\mu\nu}F_{\mu\nu}$  is 
the winding number of the gauge field configuration. For the 
instanton  configuration (\ref{phi},\ref{A}), we have $q=n$.  
\subsection{Fermionic zero modes} 
According to the index theorem (see for example \cite{B}), the Dirac 
operator in the background of the instanton satisfies the 
following relation: $\dim \ker[K]-\dim\ker[K^\dagger]=n$. 
As the instanton in 1+1 dimensions 
coincides with the vortex, these zero modes may be found by carrying out a 
similar analysis as in \cite{jackrossi}; where the 
fermionic zero modes on the Nielsen-Olesen 
string were analyzed for non chiral fermions. 
In this subsection we present the 
corresponding equations. 
 
The Lagrangian for the fermion $j$ in the background of the scalar 
and vector fields may be written as 
$\mathcal{L}^{E}_{f_j}=\overline{\Psi}^{j} K^j \Psi^j$, where  
\begin{equation} 
K^j=\left(  
\begin{array}{cc} 
-if^j \phi^* & i\partial_0-\frac{e}{2}A_0 -\partial_1-i\frac{e}{2}A_1\\ 
-i\partial_0-\frac{e}{2}A_0 -\partial_1+i\frac{e}{2}A_1 & if^j \phi  
\end{array}\right). 
\label{K} 
\end{equation} 
In the following, the family dependent 
Yukawa coupling $f^j$ will be replaced by $f$
keeping in mind that there is no mixing between different fermionic 
generations. 
 
The zero modes are the regular normalizable solutions of the equation 
$K \Psi=0$, with $A_\mu$ and $\phi$ given by (\ref{phi},\ref{A}) 
\footnote{However, in the massless case ($f^j=0$), a logarithmically 
divergent wave function is generally kept as a relevant solution. The 
reason is that its classical action is finite \cite{Ringwald}.}. Using polar
coordinates and performing the substitution
$\tilde{\Psi}=\exp\left[ \int_{0}^{r} 
\frac{A(\rho)}{2}d\rho\right]\Psi$ we get: 
\begin{equation} 
\left(  
\begin{array}{cc} 
-iFf(r)e^{-in\theta} & ie^{i\theta}\left(\frac{\partial}{\partial 
r}+\frac{i}{r}\frac{\partial}{\partial \theta}\right)\\ 
-ie^{-i\theta}\left(\frac{\partial}{\partial 
r}-\frac{i}{r}\frac{\partial}{\partial \theta}\right) & 
iFf(r)e^{in\theta}  
\end{array}\right)\tilde{\Psi}=0,\label{Kpsi} 
\end{equation} 
where $F=fv$ is the fermion mass. With the use of the phase decomposition 
$\tilde{\Psi}=\sum_{m=-\infty}^{\infty}e^{im\theta}\Psi^m$, 
equation (\ref{Kpsi}) can be rewritten as 
\begin{eqnarray} 
Ff(r)\Psi^m_L-\left(\frac{\partial}{\partial 
r}-\frac{m-n-1}{r}\right)\Psi_R^{m-n-1}&=&0,\notag\\ 
\left(\frac{\partial}{\partial 
r}+\frac{m}{r}\right)\Psi_L^{m}-Ff(r)\Psi_R^{m-n-1}&=&0. 
\end{eqnarray} 
In our case, the analysis of \cite{jackrossi} shows that for 
a vortex with topological number $n<0$ there are exactly $|n|$ fermionic 
zero modes in the spectrum of $K$ with $m$ in the interval 
$m\in \{-n+1,..,1,0\}$ and none in the spectrum of $K^\dagger$. 
For $n>0$ there are no zero modes in the spectrum of $K$, but $n$ in the
spectrum of $K^\dagger$. 

For the 
case of $n=-1$ studied below the explicit form of the zero mode is 
given by 
\begin{equation} 
\Psi^0_L(r)=\Psi^0_R(r) \propto \exp\left( -\int_0^r \left\{Ff(r')+
\frac{e}{2}A(r')\right\}dr'\right). \label{mode0}
\end{equation}
Note that for massless fermions ($F=0$), the zero mode decreases 
as $\frac{1}{\sqrt{r}}$ 
for large $r$. It is therefore not normalizable and has a divergent action. 
This behavior differ form the case of \cite{detpsim0}, \cite{Hortacsu:1979fg},
\cite{Ringwald}, where massless fermions of 
charges $e$ were considered. In their case the fermionic zero mode decreases 
as $\frac{1}{r}$ for large $r$ and has a finite action.
 
\subsection{Determinant}   
 
Due to the presence of the fermionic zero modes, 
instanton transitions imply the creation of a net number of fermions.
In the following, we will be interested in the creation of one of each type of
 fermion, for which an instanton of charge $n=-1$ is needed.
The corresponding transition probability is proportional to  $\det'K$, 
where the prime means omission of the zero eigenvalue 
in the calculation of the determinant. 

It is well known \cite{thooft} that the eigenvalue 
problem for the operator $K$ is ill defined. Consequently one has to consider 
the Laplacian type operators $K^\dagger K$ or $KK^\dagger$ which have 
the same set of eigenvalues (except for the zero modes). Then $\det' K$ is 
defined up to a phase as $\det'[K]=\det'[K^\dagger K]^{1/2}$. 
The explicit expression for the operator $K^\dagger K$ reads: 
\begin{eqnarray} 
&&K^\dagger K=\\&&\left[\begin{array}{cc} 
f^2|\phi(r)|^2-(\partial_\mu-i\frac{e}{2}A_\mu)^2+\frac{e}{2} 
\epsilon_{\mu\nu} \partial_\mu A_\nu & 
-if[\phi(eA_0+ieA_1)+(i\partial_0-\partial_1)\phi]\\ 
if[\phi^*(eA_0-ieA_1)-(i\partial_0+\partial_1)\phi^*] & 
f^2|\phi(r)|^2-(\partial_\mu+i\frac{e}{2}A_\mu)^2+\frac{e}{2} 
\epsilon_{\mu\nu} \partial_\mu A_\nu 
\end{array}\right]. \notag
\label{KdaggerK} 
\end{eqnarray} 
The fermionic equations  
 of motion, for instance equation (\ref{Kpsi}), remain 
unchanged after the variable changes (\ref{dimless}), if $f$ is replaced 
by $f/e$ and $e$ is set to $1$. The only free parameter
in the bosonic sector (\ref{actionclassiquesansdim}) is $\mu$, while 
there is a second parameter in the fermionic sector: 
 the Yukawa coupling $f$.

In conclusion, we are left with two dimensionless parameters, and the 
determinant can be calculated as a function of 
\begin{equation} 
\frac{m_H}{m_W}=\mu\sqrt{2} \quad\text{and}\quad 
\frac{F}{m_H}=\frac{f}{e}\frac{m_W}{m_H}=\frac{f}{\sqrt{2\lambda}}. 
\label{dimlessparam} 
\end{equation} 

Obviously the determinant, being a product of an infinite number of 
eigenvalues, is a divergent quantity. In the next section we discuss 
its regularization and renormalization. 
\section{Regularization and renormalization}
For perturbative calculations, the dimensional regularization 
is best suited.
However, as has been observed in \cite{thooft}, it is not 
applicable to the computation of the fermionic determinant because 
the continuation of the instanton fields to a space with 
fractional number of dimensions is not uniquely defined. 
Nevertheless, we discuss the
dimensional regularization to fix the meaning of the Lagrangian parameters
in section 3.1. 
In section 3.2, we consider another regularization 
scheme based on partial waves decomposition. It permits to 
exploit the spherical symmetry and turns out to be  
convenient for numerical purposes.
In appendix A, we consider the Pauli-Villars regularization  
used in \cite{thooft} and prove its equivalence with the partial 
waves procedure.
\subsection{Dimensional regularization}
The Lagrangian depends on
four parameters; the charge $e$, the scalar coupling $\lambda$, the 
scalar mass $m_H$, and the Yukawa coupling $f$. The model under 
consideration is super-renormalizable. In order to 
use the dimensional regularization we have to define the $\gamma$-matrices
for an arbitrary number $d=2-\varepsilon$ of dimensions:
\begin{eqnarray}
\left\{\gamma^\mu, \gamma^\nu\right\}&=&2g^{\mu\nu},\notag\\
\tr(\gamma^\mu \gamma^\nu)&=&2g^{\mu\nu},\quad  \mu,\ \nu = 0,1,...,d-1.
\end{eqnarray}
The definition of the $\gamma^5$ matrix is ambiguous, we follow here the
usual definition:
\begin{eqnarray}
\left\{\gamma^5, \gamma^\nu\right\}&=&0,\quad \nu = 0,1.\notag \\
\left[\gamma^5, \gamma^\nu\right]&=&0,\quad \nu = 2,...,d-1.
\end{eqnarray}
The physical parameters $e,~\lambda,~m_H,~f$ in two dimensions are related to 
the $d$-dimensional parameters $e_d,~\lambda_d,~m_{Hd},~f_d$ by:
$$ e_d=e\mu^{1-\frac{d}{2}}, \quad \lambda_d=\lambda \mu^{2-d}, 
\quad m_{Hd}=m_{H}\mu^{1-\frac{d}{2}}, \quad f_d=f\mu^{1-\frac{d}{2}}.$$
We will work in the $R_\xi$ gauge. The complex field 
$\phi$ is written as 
$\phi=v+h+i\varphi$, where $h$ and $\varphi$ are real. The gauge 
fixing term is
\begin{equation}
\mathcal{L}_{\mathrm{g. f.}}=\frac{1}{2}\int d^2x~G[A,h,\varphi]^2,
\end{equation}
where
\begin{equation}
G[A,h,\varphi]=\frac{1}{\xi}\left(\partial_\mu A^\mu-\xi ev\varphi\right).
\end{equation}
The Lagrangian for ghost fields $c$ is
\begin{equation}
\mathcal{L}_{\mathrm{ghost}}=\int d^2x \bar{c}\left[-\partial^2-\xi e^2 
v^2\left(1+\frac{h}{v}\right)\right]c.
\end{equation}
In the following we will work in the minimal subtraction scheme.
The only divergent parameter is the Higgs mass $m_H$.
A straightforward computation gives the relevant part of the effective action,
\begin{equation}
S^{UV}_{count}=\int d^dx \frac{1}{2}
\left(\phi^2-v^2\right)\delta m^2 \label{sdim}
\end{equation}
with
\begin{eqnarray}
\delta m^2=\frac{1}{4\pi}\left[3\lambda \left\{\ln(\frac{\mu^2}{m_H^2})+\left(\frac{1}
{\varepsilon}\right)_{\overline{MS}}\right\}+
\left(\lambda-e^2\right)\left\{ \ln(\frac{\mu^2}{m_W^2})+\left(\frac{1}
{\varepsilon}\right)_{\overline{MS}}\right\}
\right.\notag\\\left.-
2f^2\left\{\ln(\frac{\mu^2}{F^2})+\left(\frac{1}
{\varepsilon}\right)_{\overline{MS}}\right\}
\right],\label{dm}
\end{eqnarray}
where $\left(\frac{1}
{\varepsilon}\right)_{\overline{MS}}=\frac{1}{\varepsilon}-\gamma+\ln(4\pi)$.
In the minimal subtraction scheme we subtract the counterterm
$$S^{UV}_{count}=\int d^2x \frac{1}{2}
\left(\phi^2-v^2\right)\delta m_{\overline{MS}}^2,$$
with $\delta m_{\overline{MS}}^2$ containing all terms in (\ref{dm}) 
proportional to $\left(\frac{1}
{\varepsilon}\right)_{\overline{MS}}$.

For the photon propagator, the bosonic loops do not introduce any 
renormalization. However, as is well known \cite{B}, there
is a finite contribution coming from fermionic loops. Because of 
the ambiguities in the definition of $\gamma_5$, dimensional 
regularization breaks the chiral gauge invariance and a term 
$\frac{e}{4\pi}A_\mu^2$ needs to be added to the action.
The complete counterterm action to be subtracted from the initial action 
(\ref{L}) reads:
\begin{equation}
S_{count}=\left\{\int d^d x \left(-\frac{e}{4\pi}A_\mu^2+ \frac{1}{2}
\left(\phi^2-v^2\right)\delta m_{\overline{MS}}^2\right)\right\}.
\end{equation}
\subsection{Partial wave regularization}
The spherical symmetry of the instanton suggests that partial wave
expansion can be used. The eigenvalue problem decouples into  
one-dimensional differential equations. In this section we discuss a natural
way to regularize the partial waves.
We consider here only the fermionic sector.
\subsubsection{Partial wave expansion}
We may write $\det[K^\dagger K]$ as a path integral:
\begin{equation}
\det[K^\dagger K]=\int\mathcal{D}\eta\mathcal{D}\bar{\eta} \exp\left[
\int d^2x~ \bar{\eta}K^\dagger K \eta\right].\label{seta}
\end{equation}
Partial wave decomposition is defined as follows:
\begin{equation}
\eta(r,\theta)=\sum_{m=-\infty}^\infty e^{im\theta}\eta^m(r),\quad 
\bar{\eta}(r,\theta)=\sum_{m=-\infty}^\infty e^{-im\theta}\bar{\eta}^m(r).
\end{equation}
The regularization is done by putting our system in a finite spherical box of
radius $R$, and cutting the sum over the partial waves at some $m=L$. 
After performing the partial wave decomposition, the regularized
action reads:
\begin{equation}
S(R,L)=\int_0^R 2\pi r~dr\sum_{m,l=-L}^L\bar{\eta}^m(r)M^{ml}(r)\eta^l(r),
\end{equation}
with 
\begin{equation}
M^{ml}=\frac{1}{2\pi}\int d\theta e^{-im\theta}K^\dagger K e^{il\theta}.
\label{Mlm}
\end{equation}
From the general expression (\ref{KdaggerK}) for  
$K^\dagger K$, we get for the vacuum:  
\begin{equation} 
K^\dagger K_{vac}=\left[\begin{array}{cc} 
F^2-\partial_0^2-\partial_1^2 & 0\\ 
0 & F^2-\partial_0^2-\partial_1^2\end{array}\right].
\end{equation}
After phase decomposition we obtain a diagonal matrix in both spinor and 
partial wave space:
\begin{equation} 
M^{ml}_{vac}=\delta^{ml}\identity_2\left[-\frac{\partial^2}{\partial 
r^2}-\frac{1}{r}\frac{\partial}{\partial 
r}+\frac{m^2}{r^2}+F^2\right],\label{Mvac}
\end{equation}
where $\identity_2$ is the identity in spinor space.
The radial eigenvalue equation in vacuum reads:
\begin{equation} M^{mm}_{vac}\eta^m_\lambda=\lambda^2\eta^m_\lambda,
\end{equation}
with boundary conditions 
\begin{eqnarray} 
&&\eta^m(0)=\eta^m(R)=0,\quad m\neq 0,\notag\\
&&\eta^0(0)=1,\quad\eta^0(R)=0.\label{bc}
\end{eqnarray}
From the relations (\ref{Mvac})-(\ref{bc}) the free propagator
may be derived
\begin{eqnarray} 
G^R_m(r,r')&=&\sum_\lambda \frac{\bar{\eta}^m_\lambda(r)\eta^m_\lambda(r')}
{\lambda^2}\notag
\\&=&\frac{\identity}{2\pi}\left\{\begin{array}{cc} 
\frac{I_m(Fr)}{I_m(FR)}\left[K_m(Fr')I_m(FR)-I_m(Fr')K_m(FR)\right],& 
\quad r<r',\\ 
\frac{I_m(Fr')}{I_m(FR)}\left[K_m(Fr)I_m(FR)-I_m(Fr)K_m(FR)\right], & 
\quad r>r'.\end{array}\right.\label{Gm} 
\end{eqnarray}
It allows us to treat the interaction terms present in (\ref{KdaggerK}) 
by standard diagrammatic methods. 
\subsubsection{One-loop divergences in partial waves} 
As we have already seen, the fermionic parameter $f$
needs no renormalization. However, the mass of the scalar Higgs 
receives divergent contributions from fermionic diagrams.
The partial wave regularization can't be introduced at the level of
 the fermionic Lagrangian (\ref{L}), but only at the level of the 
squared determinant (\ref{seta}). One does not expect that the 
counterterms derived from the initial Lagrangian are sufficient to 
remove all infinities in (\ref{seta}). 
Hence, we recalculate the counterterm action
(see appendix B for details) needed to renormalize (\ref{seta}). 
The result is:
\begin{eqnarray} 
S_{count}^{UV}(L,R)&=&\sum_{m=-L}^L S^m_{count}(R)\notag\\&=&\sum_{m=-L}^L
\int_0^R 2\pi r 
~\tr[G^m(r,r)]\left(f^2\left(|\phi|^2-v^2\right)+\frac{e}{2}
\varepsilon_{\mu\nu}\partial_\mu 
A_\nu\right)dr\label{diagpv}, 
\end{eqnarray} 
where the $S^m_{count}$ are finite for each $m$ and only the sum is 
divergent in the limit $L\to\infty$. Note that the 
counterterm (\ref{diagpv}) is non-local. This is due to the 
non-locality of the partial wave regularization 
procedure and may be checked to be correct by comparison to Pauli-Villars 
regularization, see appendix A.6.

For small constant background fields, (\ref{diagpv}) leads to
\begin{eqnarray} 
S_{count}^{UV}&=&\int_0^R\left(f^2\left(|\phi(r)|^2-v^2\right)
+\frac{e}{2}\varepsilon_{\mu\nu} 
\partial_\mu A_\nu(r)\right)d^2r\frac{1}{2\pi} 
\left[1+\log\left(\frac{4L^2}{F^2R^2}\right)\right]\label{divpv}. 
\end{eqnarray}
In order to get results in the $\overline{MS}$-scheme from those 
calculated in the partial waves, we calculate the difference 
$\delta S^{UV}_{count}$ between the effective action found in these two 
schemes\footnote{
The counterterms found in the dimensional regularization have to be
 multiplied by a factor of 2 because we are dealing here with the 
squared operator $K^\dagger K$.}. 
The result reads
\begin{eqnarray}
\delta S^{UV}_{count}=\left(\log\left(\frac{4 L^2}{R^2\mu^2}\right)-
\left(\frac{1}{\varepsilon}\right)_{\overline{MS}}\right)\int
\frac{d^2x}{2\pi}f^2(|\phi|^2-v^2)+S_{gf},\label{46}
\end{eqnarray}
where
\begin{equation}
S_{gf}= \log\left(\frac{4 L^2}{R^2F^2}
\right)\int\frac{d^2x}{2\pi}\frac{e}{2}\epsilon_{\nu\rho}\partial_\nu 
A_\rho(x).\label{sgf}
\end{equation}

In comparison with the dimensional regularization, a 
supplementary divergent term involving gauge fields $S_{gf}$ has appeared. 
It also arises when using Pauli-Villars 
scheme (see appendix A.3) and is an extra divergence of the action 
(\ref{seta}) in comparison to the initial action (\ref{L}).
If we Wick-rotate $S_{gf}$ back to our initial Lagrangian  
in Minkowski space-time, 
it gets an extra factor of $i$, the action becomes non-hermitian and breaks
unitarity.
Because of this, $S_{gf}$ must be subtracted completely.

For the photon propagator, as in dimensional regularization, 
we have to subtract from the effective action the term
\begin{equation} 
S_{count}^{IR}(R)=\frac{e^2}{4\pi}\int_0^R d^2r A^2(r)
\label{Scontinfra} 
\end{equation} 
to recover chiral gauge invariance (see appendix B.1).
\subsubsection{Regularization and renormalization in partial waves} 
From the counterterm (\ref{divpv}), we see that
the initial theory is recovered in the limit $\frac{2L}{FR}\to \infty$.
The summation over partial waves and the limit $L\to\infty$ 
has to be performed first and the infinite volume limit must be 
taken only after having removed the infrared counterterm.  
  
The explicit expression for the counterterms in the case of 
space dependent background is obtained in 
integrating (\ref{diagpv}) and the renormalized fermionic determinant may 
formally be written as: 
\begin{eqnarray} 
&&\det_{ren}[K^\dagger K]=\label{detrenpw}\\&&\lim_{R\to\infty}\left( \lim_{L\to\infty}\left[
\prod_{m=-L}^{L}\frac{\det[M^m_{inst}]}{\det[M^m_{vac}]}\exp\left\{
-S_{count}^{UV}(L,R)\right\}\right]\exp\left\{-S_{count}^{IR}(R)\right\}
\right).\notag
\end{eqnarray}
This prescription differs from the one of \cite{BK} where the 
limit $R\to \infty$ is taken first. It is shown in Appendix C that
the order of the limits is crucial.
\section{Determinant calculation} 
After the partial wave decomposition (\ref{seta}-\ref{Mlm}),
 $K^\dagger K(r,\theta)$ was expressed in terms of $M^{lm}(r)$. 
For our purposes, the
 case where $M^{lm}$ is diagonal in partial wave space 
($M^{lm}=\delta^m_l M^{m}$) is sufficient\footnote{We are mainly interested 
in the case $n=-1$, where one of each type of 
fermions is created. In this particular case, the operator $M^{ml}_{inst}$
 is diagonal in partial wave space. Note that this point is not crucial,
as explained in section 4.1, the determinant may be calculated 
in the non-diagonal case as well.}. The determinant may be calculated as:
\begin{equation} 
\det[K^\dagger K]=\prod_{m=-\infty}^{+\infty}\det[M^{m}].\label{detMm} 
\end{equation} 
We are left with the much easier problem of finding the 
determinant of one-dimensional operators, which may be addressed 
with the following theorem \cite{coleman}: Let us consider two operators 
$O_i=-\partial_x^2+W_i(x),~i=1,2$ defined in an interval of length $R$. 
Let $\Psi_i,~i=1,2 $ be the solution of $O_i \Psi_i=0$ with the boundary 
conditions 
\begin{equation}\Psi_i(0)=0,\Psi_i'(0)=1,~i=1,2, \label{cb}\end{equation}
 we have:
\begin{equation}
\det\left[\frac{O_1}{O_2}\right]=\frac{\Psi_1(R)}{\Psi_2(R)} \label{thmdet}. 
\end{equation} 
\subsection{Treatment of radial operators} 
We follow here the method developed in \cite{BK} to calculate 
determinants. Note that here we will first consider the radial problem 
for $0$ to $R$ where $R\gg 1$ and the limit $R\to\infty$ will 
be taken afterward.

In the present case, even if $M^{ml}$ is diagonal in partial wave space, it is
not diagonal in spinor space. The theorem (\ref{thmdet}) 
needs generalization to two
coupled second order differential equations.
We are interested in the ratio between the operator
$$M^m=\left(\begin{array}{cc} M_{11}^m & M_{12}^m\\ M_{21}^m & 
M_{22}^m\end{array}\right)$$ in the instanton 
background and the vacuum operator $M^{m,vac}$, which is assumed to be diagonal. 
Let us define the matrix $\psi^m_{ij}$ ($i,j=1,2$) and $\psi^{m,vac}_{L,R}$
 as the solutions of the following differential systems: 
\begin{equation}\begin{array}{cc} 
\sum_j M^m_{ij}\psi^m_{j1}=0, & M^{n,vac}_{11}\Psi^{m,vac}_L=0,\\ 
\sum_j M_{ij}\psi^m_{j2}=0, & M^{n,vac}_{22}\Psi^{m,vac}_R=0, \label{Mpsi}
\end{array}\end{equation} 
with boundary conditions
\begin{eqnarray} 
&&\lim_{r\rightarrow 
0}\frac{\psi_{11}^m}{\psi^{vac,m}_L}=1, \quad \lim_{r\rightarrow 
0}\frac{\psi_{21}^m}{\psi^{vac,m}_R}=0,\notag\\  
 &&\lim_{r\rightarrow 0}\frac{\psi_{12}^m}{\psi^{vac,m}_L}=0, \quad 
\lim_{r\rightarrow 0}\frac{\psi_{22}^m}{\psi^{vac,m}_R}=1. \label{Mbc}
\end{eqnarray} 
The determinant is then given by: 
\begin{equation} 
\frac{\det[M^m]}{\det[M^{m,vac}]}=\frac{\det[ 
\psi_{ij}^m(R)]}{\psi^{m,vac}_L(R)\psi^{m,vac}_R(R)}.\label{detC} 
\end{equation} 
The remaining determinant is just the usual determinant for 
$2\times2$ matrices. It is an easy exercise to reproduce step by step 
the demonstration of \cite{coleman} in this more general case. 
 
The vacuum operator which is given in (\ref{Mvac}), has an analytic 
solution $\Psi^{m,vac}_j=I_m(Fr)$.
For the instanton ($n=-1$) configuration we get: 
\begin{eqnarray} 
M_{11}^m&=&-\frac{\partial^2}{\partial 
r^2}-\frac{1}{r}\frac{\partial}{\partial 
r}+\frac{m^2}{r^2}+F^2f^2(r)+\frac{e}{2}\varepsilon_{\mu\nu}\partial_\mu A_\nu
+ \frac{e^2}{4}A^2(r) + me\frac{A(r)}{r}, \notag\\
M_{12}^m&=&M_{21}^m=
F\left(-f'(r)-\frac{1}{r}f(r)+eA(r)f(r)\right),\notag\\ 
M_{22}&=&  -\frac{\partial^2}{\partial 
r^2}-\frac{1}{r}\frac{\partial}{\partial 
r}+\frac{m^2}{r^2}+F^2f^2(r)+\frac{e}{2}\varepsilon_{\mu\nu}\partial_\mu A_\nu
+ \frac{e^2}{4}A^2(r)- me\frac{A(r)}{r}. \label{Minst}  
\end{eqnarray} 
The solution $\Psi_{ij}^m(r)$ needs to be computed numerically. 
To this aim, it is convenient to make the following 
substitution: 
\begin{equation} 
\Psi_{ij}^m(r)=\left(\delta_{ij}+h_{ij}^m(r)\right)I_m(r), 
\end{equation} 
The determinant is then evaluated with
\begin{equation} 
\frac{\det[M^{m}]}{\det[M^{vac,m}]}=\det[\delta_{ij}+h^m_{ij}(R)]. 
\end{equation} 
In terms of 
the functions $h^m_{ij}$, the equation (\ref{Mpsi}) takes the form 
of an ordinary
quantum mechanical equation with potential $V_{ij}(r)$:
\begin{equation} 
\left[\frac{\partial^2}{\partial r^2}+\left(\frac{1}{r} + 
2\frac{I_m'(Fr)}{I_m(Fr)}\right)\frac{\partial}{\partial 
r}\right]h^m_{ij}(r)=V_{ik}(r)\left(\delta_{kj}+h^m_{kj}(r)\right).
\label{[]h} 
\end{equation} 
The effective potential $V_{ij}(r)$ in the background of the instanton 
is given by the following expressions:
\begin{eqnarray} 
V_{11}(r)&=&F^2\left(f^2(r)-1\right)-e\frac{A(r)}{2r}+ 
e^2\frac{A^2(r)}{4}-e\frac{A'(r)}{2}-me\frac{A(r)}{r},\notag\\ 
V_{12}(r)&=&V_{21}(r) = eFf(r)A(r)- Ff'(r)- \frac{Ff(r)}{r},\\ 
V_{22}(r)&=&F^2\left(f^2(r)-1\right)-e\frac{A(r)}{2r}+ 
e^2\frac{A^2(r)}{4}-e\frac{A'(r)}{2}+me\frac{A(r)}{r}.\notag 
\end{eqnarray} 
The functions $h_{ij}^m(r)$ can easily be found numerically from
(\ref{[]h}) with the boundary conditions
\begin{equation} 
\begin{array}{cc} h_{ij}(0)=0, & i,j=1,2,\\h'_{ij}(0)=0, & 
i,j=1,2. 
\end{array} 
\end{equation} 

For $m=0$ we have to remove the zero-mode present in $M^{0}_{inst}$. 
In this case, it is possible to diagonalize the operator $M^{m}_{inst}$ with
the substitution $\Psi_\pm=\Psi_L\pm\Psi_R$: 
\begin{eqnarray} 
&&\left[-\frac{\partial^2}{\partial 
r^2}-\frac{1}{r}\frac{\partial}{\partial r}+F^2f^2(r)
+\frac{e}{2}\varepsilon_{\mu\nu}\partial_{\mu} A_\nu 
+\frac{e^2}{4}A^2(r)\right.\notag\\ &&\pm 
\left.F\left(-f'(r)-\frac{1}{r}f(r)+eA(r)f(r)\right) 
\right]\Psi_\pm=M^0_\pm\Psi_\pm=0.\label{equamodezero} 
\end{eqnarray}
The fermionic zero-mode is contained in $M_+^0$. We calculate 
$\det[M^0_-]$ with (\ref{thmdet}) and $\det'[M^0_+]$ as in \cite{BK}: 
\begin{equation} 
\frac{\det'[M^{0,inst}_+]}{\det[M^{0,vac}_+]}=\frac{\frac{d}{d\lambda^2} 
\det[M^{0,inst}_++\lambda^2]|_{\lambda^2=0}}{\det[M^{0,vac}_+]} 
\label{modezero}=\frac{d}{d\lambda^2}h^\lambda (R).
\end{equation} 
In the last relation $h^\lambda(r)$ is defined through 
$\Psi^{inst}_{+\lambda}=\Psi^{vac}(1+h^\lambda)$ with 
$\Psi^{inst}_{+\lambda}(r)$ being a solution of $$(M^{0,inst}_++\lambda^2)
\Psi^{inst}_{+\lambda}(r)=0,$$ with the
boundary condition (\ref{cb}) and $\Psi^{vac}(r)=I_0(Fr)$ being the solution 
of $$M^{0,vac}\Psi^{vac}=0.$$
\subsection{Ultraviolet divergences} 
%
A possible way to calculate the counterterms $S_{count}^m(R)$
 is given in (\ref{diagpv}). 
We need to integrate numerically the 
Green's function multiplied by the potential 
$U(r)=f^2(|\phi|^2-v^2)+\frac{e}{2}\varepsilon_{\mu\nu}\partial_\mu A_\nu$. 
As the Green's function is not smooth, for numerical calculations, it
is more convenient to solve the related 
differential equation\footnote{A complete derivation is found in \cite{BK}, the extra factor of 2 in front of $U(r)$ comes from the trace in spinor 
space.}
\begin{equation} 
\left[\frac{\partial^2}{\partial r^2}+\left(\frac{1}{r} + 
2\frac{I_m'(Fr)}{I_m(Fr)}\right)\frac{\partial}{\partial 
r}\right]S_{count}^m(r)=2U(r)\label{[]hk(1)}, 
\end{equation} 
with the boundary conditions $S_{eff}^m(0)=0,~S_{eff}^{'m}(0)=0$. 
For the instanton configuration we have: 
\begin{eqnarray} 
U(r)\overset{inst.}=&F^2(f^2(r)-1)-\frac{e}{2}(A'(r)+\frac{A(r)}{r}). 
\end{eqnarray} 
\section{Numerical procedures} 
In this section we describe the numerical methods used in this 
work. First the background, namely the well known Nielsen-Olesen vortex  
is considered. The method used here to find the profile is explained briefly.
In the second part, the calculations related to the fermionic 
determinant are discussed, namely the integration of the  
differential equations, asymptotic solutions, subtraction of divergences 
and treatment of zero-modes. The renormalization and convergence of the  
different limits are checked and finally, results for the determinant are given. 
\subsection{Background} 
\begin{figure} 
\begin{center} 
\includegraphics[width=120mm,height=72mm]{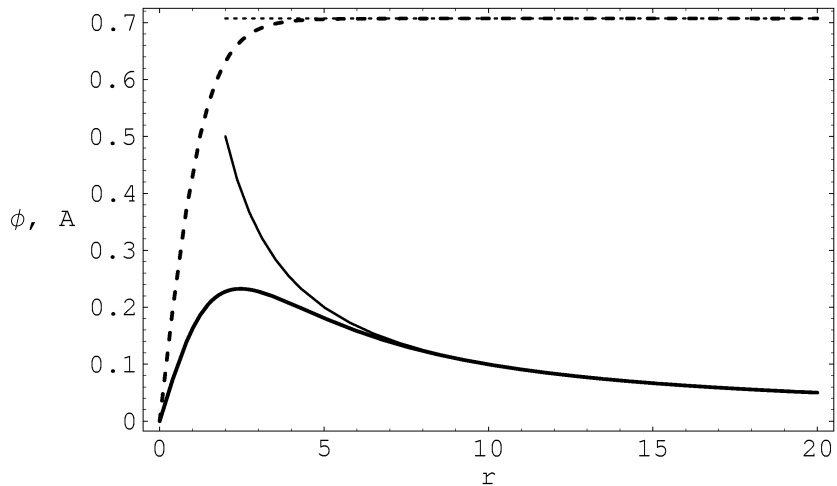} 
\caption{Instanton profile for $m_H/m_W=2$, classical fields 
$\phi(r)$ (dashed), $A(r)$ and their asymptotic forms (thin lines).} 
\label{f1} 
\end{center} 
\end{figure}

The instanton profile may be found with a shooting method (see for instance
\cite{NR}). 
The boundary conditions at $r=0$ are of the form: 
\begin{equation} 
A(0)=0 ,\quad A'(0)=b, \quad\quad \phi(0)=0 ,\quad \phi'(0)=\beta, 
\end{equation} 
where the parameters $b$ and $\beta$ are found imposing the limits 
(\ref{l2}) and (\ref{l4}).
We start the numerical integration at $r \sim 10^{-7}$ instead 
of $r=0$, where some trivial divergences occur, and use a small $r$ expansion 
for $\phi$ and $A$: 
\begin{eqnarray} 
A(r)&=&b r-\frac{\beta^2}{8}r^3 + \mathcal{O}(r^5),\notag\\ 
\phi(r)&=&\beta r-\frac{\beta (1+2b)}{8}r^3 + \mathcal{O}(r^5),
\end{eqnarray}
valid for $n=-1$.
The 
numerical integration is done with 32 decimals, and to get an 
accurate\footnote{ 
The accuracy can be checked by calculating the instanton number 
(\ref{topcharge}), or the action of the instanton for 
$\frac{m_H^2}{m_W^2}=1$ that is known to be $\pi v^2$ 
\cite{cordean}. The results of the numerical integration agrees to 13 
decimals with the action in the latter case and at least 7 for the 
instanton number in any case (see figure \ref{f1} and table 
\ref{table1}).} profile, the boundary conditions have to be specified within 
an accuracy of order $\sim 10^{-14}$. 
%
\subsection{Fermionic determinant} 
For the fermionic determinant, the task is to solve equations 
(\ref{[]h}) and (\ref{[]hk(1)}). These equations are 
completely symmetrical under the change $m\rightarrow-m$ therefore 
only positive $m$ need to be considered. As the solution to the fermionic 
equations in the asymptotic instanton fields is known, it is sufficient 
to integrate numerically to $r\sim 15$ and glue the asymptotic solution 
\begin{eqnarray} 
\Psi_L^m(r)=A I_{m-1/2}(Fr)+B K_{m-1/2}(Fr),\notag \\ \Psi_R^m(r) =C 
I_{m+1/2}(Fr)+D K_{m+1/2}(Fr). 
\end{eqnarray} 
The constants $A,~B,~C,~D$ are determined in imposing 
the continuity of $\Psi^m(r)$ and its first derivative. The 
numerical integration, like for the vortex, starts at $\epsilon 
\sim 10^{-6}$, where the boundary conditions are found by calculating the 
power expansion for the $h^m_{ij}$:
\begin{eqnarray} 
h^m_{ij}(\epsilon)=V_{ij}(\epsilon)\frac{\epsilon^2\delta_{ij}}{2(2+2m)}, 
\quad 
h'^m_{ij}(\epsilon)=V_{ij}(\epsilon)\frac{\epsilon\delta_{ij}}{(2+2m)},\notag\\
S^m_{count}(\epsilon)=U(\epsilon)\frac{\epsilon^2}{2(2+2m)}, 
\quad 
S'^m_{count}(\epsilon)=U(\epsilon)\frac{\epsilon}{(2+2m)}. 
\end{eqnarray} 
Having found the $h^m_{ij}$, we calculate the partial determinants 
with (\ref{detC})
and subtract to each wave the partial counterterm $S^m_{count}$ found with 
(\ref{[]hk(1)}) as prescribed in (\ref{detrenpw}).

Numerically we store the value of the determinant for $\sim 50$ different 
values of system radius $R_i,~i=1,...,50$. 
\begin{figure} 
\begin{center} 
\includegraphics[width=100mm,height=60mm]{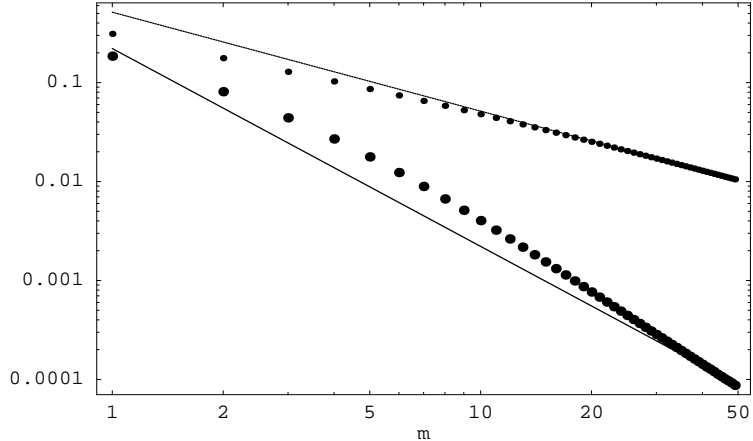} 
\caption{Logarithm of the partial determinant $\log(\det[M^m])$ as 
a function of angular momentum $m$, for the parameter values 
$\frac{m_H^2}{m_W^2}=1$, $\frac{F}{m_H}=0.1$. The upper small dots shows 
the results for determinant before renormalization, and the lower 
ones after. The lines show respectively $1/m$ and $1/m^2$ 
behavior.} 
\label{f2} 
\end{center} 
\end{figure} 
After renormalization, the partial determinants $\det[M^m]$ 
have to decrease at least as $\frac{1}{m^2}$, so that the 
product over $m$ remains finite. This is checked in figure 2. 
Using this property, for each $R_i$, we calculate the partial
determinants from $m=1$ to $m=L\sim 30$ and fit   
them with an inverse power law: 
$$\det_{ren}[M^m]= 
\frac{const_2}{m^2}+\frac{const_3}{m^3}+ 
\frac{const_4}{m^4}.$$
This approximate expression is then used for  
$m=L\sim 30$ to infinity. 

This completes the limit $L\to\infty$ and we may
consider the limit $R\to \infty$. At this point the determinant still 
depends on
$R$ (see figure 3), according to (\ref{detrenpw}), the infrared 
counterterm (\ref{Scontinfra}) have to be subtracted. The renormalized 
determinant becomes approximately
constant for typically $10<FR<100$ and $FR$ can be chosen in this range. 
Keeping in mind that for large $FR$ higher partial wave should be considered,
it is expected that the result becomes inaccurate at large $FR$. 
Fortunately the determinant converges very fast as $FR\to\infty$ and is found  
to be constant up to 4 decimals for typically
 $20<FR<40$ from were the result is extracted. 
\begin{figure} 
\begin{center} 
\includegraphics[width=100mm,height=60mm]{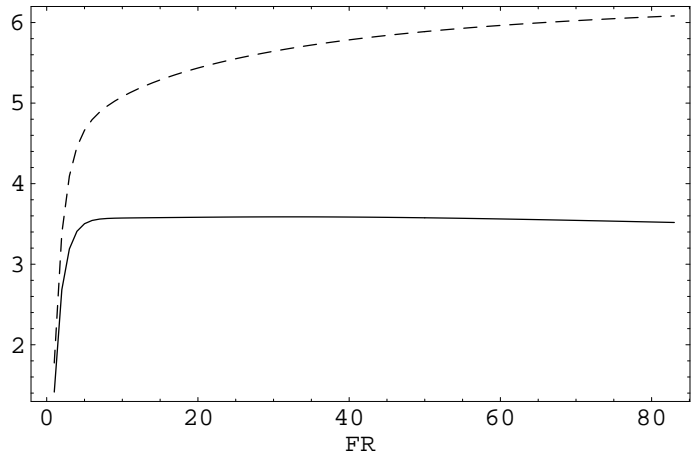} 
\caption{Logarithm of the determinant $\log(\det[M])$ as function of 
the system radius R, before (dashed) and after subtracting the infrared 
counterterm (\ref{Scontinfra}) ($\frac{m_H^2}{m_W^2}=1$, 
$\frac{F}{m_H}=0.1$)} 
\label{f3} 
\end{center} 
\end{figure} 

For $m=0$ the zero-mode in $M^0_+$ has to be removed in the 
determinant calculation. This is done with(\ref{modezero}),
 where the derivative is approximated as
\begin{equation} 
\det'[M^0_+]=\frac{d}{d\lambda^2}h^\lambda(R)\cong 
\frac{h^\lambda(R)-h^0(R)}{\lambda}.\label{detmodezero} 
\end{equation} 
To get an accurate result, we take $\lambda^2$ of the order 
$(10^{-3})F^2$ and perform the computation of $h^\lambda(R)$
 for some $(\sim 10)$ different 
values of $\lambda$. These results are fitted to extrapolate 
the value of (\ref{detmodezero}) at $\lambda=0$.
\subsection{Results}
\begin{figure} 
\begin{center} 
\includegraphics[width=120mm,height=70mm]{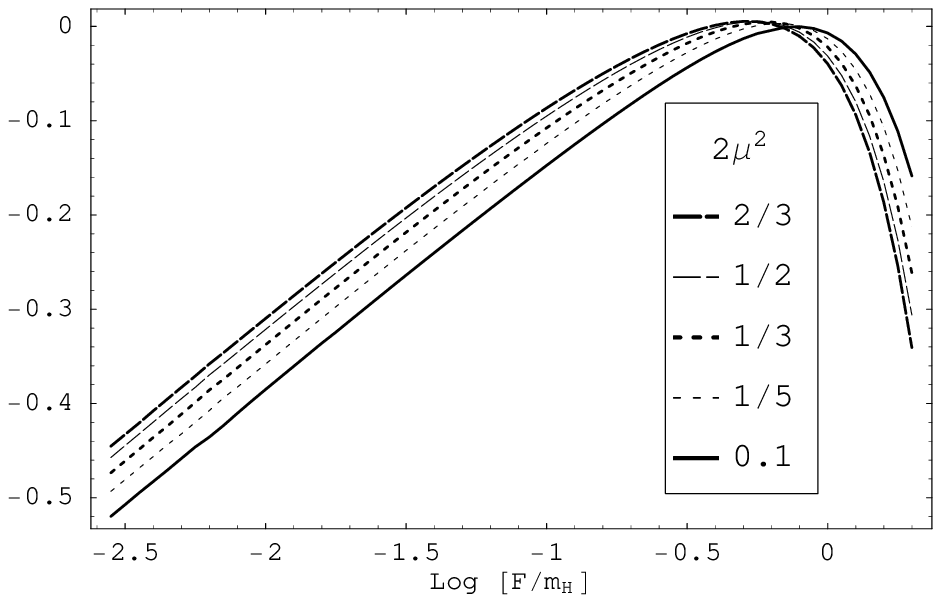} 
\includegraphics[width=120mm,height=70mm]{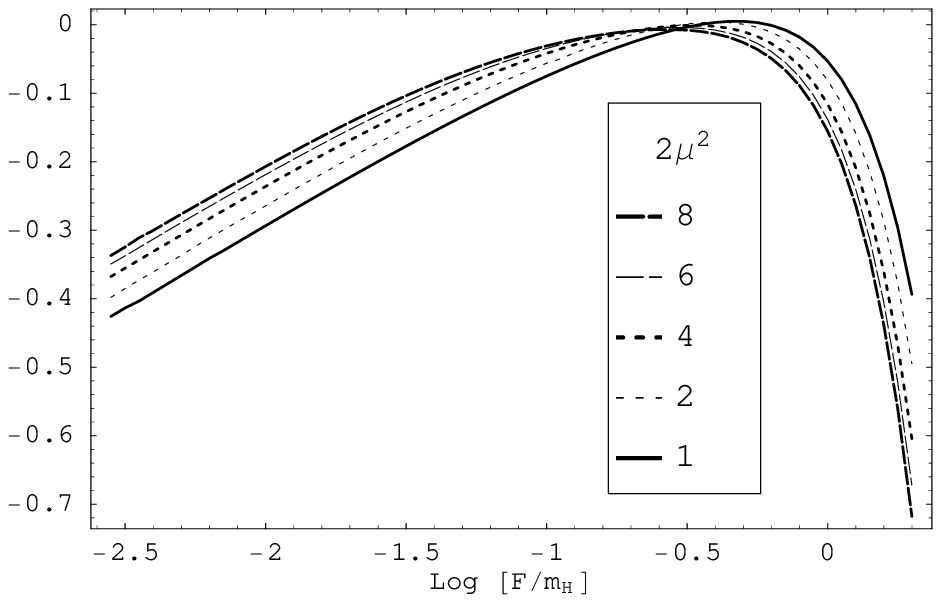} 
\caption{Logarithm of the determinant $\log[F\sqrt{\det'[K^\dagger K]}]$ 
as a function of the 
dimensionless fermion mass $\frac{F}{m_H}$ (horizontal axis) for different
values of $2\mu^2=\frac{m_H^2}{m_W^2}$. The different values 
of the determinant 
are fitted to few percents accuracy with the following expression:
$ 
F\sqrt{\det'[K^\dagger K]}=1.62\mu^{1/10}F^{1/4}+\left(-4.60+3.71\mu^{1/5}-0.632\ln\mu\right)F
+\left( 6.97-6.76\mu^{1/5}+0.866\ln\mu\right)\ln(1+F) 
$ in the interval $0\leq F\leq 3$ and $0.05\leq\mu^2\leq 8$.}
\label{ff}
\end{center}
\end{figure}

We first note that $\det'[M]$ has dimension of $\textrm{mass}^{-2}$ from 
$\frac{d}{d\lambda^2}$ in (\ref{modezero}). The fermion mass $F$ may be used 
to obtain a dimensionless 
quantity  $F^2\frac{d}{d\lambda^2}\det[M]$. The 
results for $F\sqrt{\det'[K^\dagger K]}$ are plotted in figure 
\ref{ff}. 
The logarithm of the partial determinant 
$\log(\det[M^m])$ behaves as $\frac{1}{m}$ and after renormalization 
as $\frac{1}{m^2}$, see figure \ref{f2}. It becomes constant at 
large $R$ after subtraction of the infrared counterterm, see figure \ref{f3}.  
 
\begin{figure} 
\begin{center} 
\includegraphics[width=100mm,height=70mm]{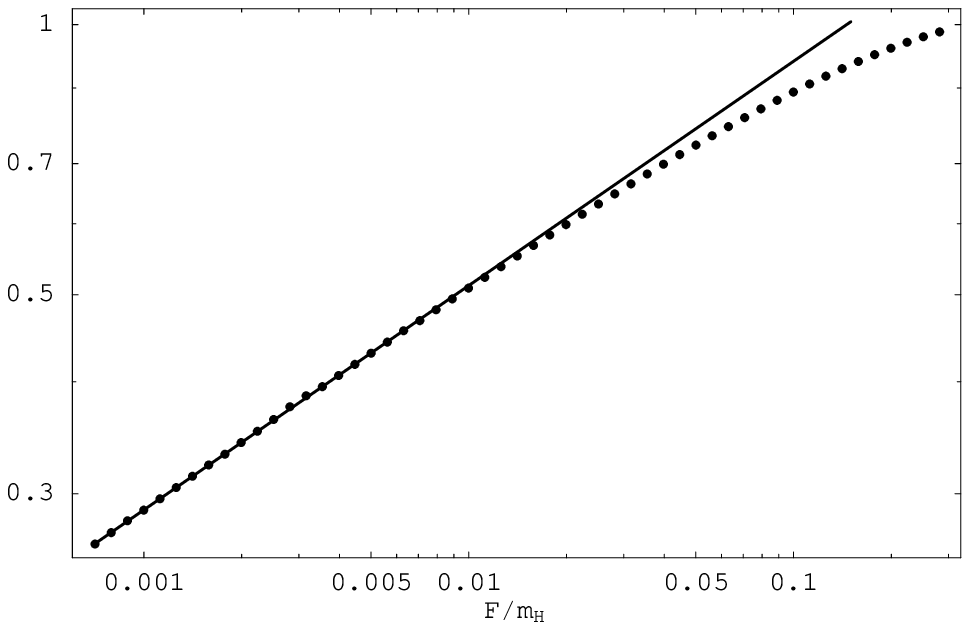} 
\caption{Determinant $F\sqrt{\det'[K^\dagger K]}$ as a function of the 
dimensionless fermion mass $\frac{F}{m_H}$ for the case 
$\frac{M_H}{m_W}=1$. The values are fitted with the power law: 
$F\sqrt{\det{M^0_+}}=1.644 \left(\frac{F}{m_H}\right)^{1/4}$.} 
\label{f5} 
\end{center} 
\end{figure}

The behavior of the 
determinant for small fermion mass $F$ is a power law, see figure 
\ref{f5}. %
This comes from the partial determinant $\det'(M^0_+)$ where 
we remove the zero mode and can be checked with some 
analytical approximation (see appendix D). The accuracy of the 
value for the determinant is estimated to be of the order $10^{-3}$ 
but may be less for $\frac{F}{m_H}<10^{-2}$.
\begin{table} 
\begin{center}
\begin{tabular}{|c|c|c|c|} 
\hline 
$\frac{m_H^2}{m_W^2}$ & $\frac{S_{cl}}{\pi v^2}$ & 
$b$ & $\beta$\\ 
\hline\hline 
1/10&0.6388286986270& 2.557798983491183&5.756251019029544\\ 
\hline 
1/5&0.7259086109970&1.554461598144364&3.541849174468259\\ 
\hline 
1/3&0.8008642959782&1.081478368993385&2.496453159112955\\ 
\hline 
1/2&0.8679102902678&0.812560321222651&1.901012558603257\\ 
\hline 
2/3&0.9199259150759&0.663981767654766&1.571374124589507\\ 
\hline 
1&1.0000000000000& 0.499999999999919& 1.206575709162995\\ 
\hline  
2&1.1567609413307&0.308286653343485&0.777359529040461\\ 
\hline 
4&1.3405945494178&0.189926436282935&0.508674018585679\\ 
\hline 
6&1.4612151896139&0.142825844043109&0.399789567459296\\ 
\hline 
8&1.5526758357349&0.116536242666195&0.338046791533589\\ 
\hline 
\end{tabular}
\caption{Results for different $\frac{m_H^2}{m_W^2}$: classical 
action ($S_{cl}$), logarithm of the fermionic determinant 
 for dimensionless fermion coupling $g=0.1$ ($\det_{F=0.1}$) and the 
boundary conditions at $r=0$ for the instanton profile ($b$ and 
$\beta$).}  
\end{center}
\label{table1} 
\end{table} 
\newpage
\section{Conclusion and outlook}  
In this paper, we have studied an instanton transition in the 
chiral Abelian Higgs model with fermion number violation and
 computed the fermionic determinant taking into account 
the Yukawa couplings. 

The dimensional regularization has been used to
fix the meaning of the Lagrangian parameters. The numerical calculations 
have been performed in the
partial wave scheme, and the Pauli-Villars regularization is studied
for completeness in Appendix A. 

In the limit of massless fermion ($F\to 0$), our results can't be compared 
to the
calculation of \cite{detpsim0}, \cite{Hortacsu:1979fg},
\cite{Ringwald}. Fermions of electric charge equal to the scalar field 
charge $e$ where considered in these previous references, whereas we 
considered pairs of fermions with half-integer charge $\frac{e}{2}$. 
The instanton transition probability
vanishes as $F^{1/4}$ in our case whereas it is finite in the case of integer
fermionic charges. As noted in section 2.3, there is no fermionic zero mode 
in our case if the fermion mass is set to zero. 
It is therefore not possible to create 
massless fermions with an instanton of charge $n=-1$. 
The fact that the probability to create fermions vanishes in the massless limit
 confirms this observation.

As can be seen in (\ref{topcharge}), considering only one family 
of fermions leads to the creation of one single fermion. 
This process seems to 
be possible in two dimensions although is it forbidden in four 
dimensions because of the Witten anomaly \cite{Witten}. 
This is an important feature of this model, 
which is addressed in \cite{future}. 
\section{Acknowledgments}
We thank F. Bezrukov, S. Khlebnikov, V. Rubakov and P. Tinyakov for helpful
discussions. This work has been supported by the Swiss Science Foundation.
\appendix 
\section{Appendix: Pauli-Villars regularization}
We compare here the Pauli-Villars regularization of \cite{thooft} to
the $\overline{MS}$ regularization and partial wave regularization.
The partial wave regularization shows some unusual features 
such as non-locality, see equation (\ref{diagpv}), and a
renormalization of the gauge field action, see equations 
(\ref{46}, \ref{sgf}). In order
to understand better their origin, let us compare the partial wave and 
the well known Pauli-Villars procedure.
In Pauli-Villars regularization, a determinant can be calculated
as in \cite{thooft}: 
\begin{equation} 
\det_{reg}[K^\dagger K_{A,\phi}]=\frac{\det[K^\dagger K_{A,\phi}]}
{\det[K^\dagger 
K_{0}]}\frac{\det[K^\dagger K_{0} + M^2]}{\det[K^\dagger 
K_{A,\phi}+M^2]},\label{regM} 
\end{equation} 
where $K^\dagger K_{A,\phi}$ and $K^\dagger K_{0}$ are respectively
 the fermionic operators (\ref{KdaggerK}) in the background of the 
fields $(A,\phi)$ and in the vacuum.
In order to determine all necessary counterterms
in this regularization scheme, one can consider small 
perturbations around the vacuum.
In principle, the instanton determinant under consideration
 may have been calculated within Pauli-Villars regularization. However, 
the partial wave analysis is technically simpler for numerical computations.
\subsection{Effective action in Pauli-Villars regularization}  
The potentially divergent terms may be extracted in calculating the 
first and second order terms in the Taylor development of the 
logarithm of (\ref{regM}) 
with respect to the fields. 
In the sections A.2, A.3, A.4, we calculate  
all the relevant functional derivatives and find their contribution to 
the determinant. The result is the following effective action:  
\begin{eqnarray} 
S_{count}^{UV} &=&\log\left(\frac{M^2}{F^2}\right)\int 
\frac{d^2x}{2\pi}\left\{\frac{e}{2}\epsilon_{\mu\sigma}\partial_\mu 
A_\sigma(x)+f^2\left(|\phi(x)|^2-v^2\right) \right\} 
.\label{diagPVi} 
\end{eqnarray}
Note that the Pauli-Villars regularization is gauge invariant
, however as in the partial waves (\ref{sgf}),
a new divergent term proportional to $\epsilon_{\mu\sigma}\partial_\mu 
A_\sigma$ arises.

To make the link between Pauli-Villars and dimensional regularization in the
minimal subtraction scheme, we may calculate the difference 
$\delta S^{UV}_{count}$ between the effective actions (\ref{diagPVi}) 
and (\ref{sdim}). 
This provides us with a way to interpret the 
Pauli-Villars parameter $M$ in terms of the parameter $\mu$ coming from 
dimensional regularization:
\begin{eqnarray}
\delta S^{UV}_{count}(\mu,M)&=&\left(\log\left(\frac{M^2}{\mu^2}\right)-
\left(\frac{1}{\varepsilon}\right)_{\overline{MS}}\right)\int 
\frac{d^2x}{2\pi} f^2\left(|\phi(x)|^2-v^2\right)\notag\\&&+\log\left(
\frac{M^2}{F^2}\right)\int 
\frac{d^2x}{2\pi} \frac{e}{2}\epsilon_{\nu\sigma}\partial_\nu 
A_\sigma(x).
\end{eqnarray}
The renormalized fermionic determinant $\det_{ren}[K^\dagger K]$ 
may be written as:
\begin{equation}
\det_{ren}[K^\dagger K]=\lim_{M\to\infty}\left(\det_{reg}[K^\dagger K]\exp
\left\{-S_{count}^{UV}\right\}\right).\label{35bis}
\end{equation}
The counterterms calculated above can be checked to be sufficient,  
in calculating 
determinants of configurations of $A$ and $\phi$ that contains small 
perturbations around the vacuum. This can be done analytically, see 
appendix A.5. 
\subsection{Functional derivatives with respect to the scalar field} 
We consider first the contributions that lead to the renormalization of the 
scalar field mass. The corresponding divergent terms can be found in calculating 
 the first and second derivatives of (\ref{regM}) with respect to the scalar field. 
The first derivatives read: 
\begin{eqnarray} 
&& \left.\frac{\delta}{\delta\phi(k)}\log\left(\det_{reg} [K^\dagger 
K_{A,\phi}]\right)\right|_{A=0,\phi=v}=\left. \notag
\frac{\delta}{\delta\phi^\dagger(k)}\log\left(\det_{reg}[K^\dagger 
K_{A,\phi}]\right)\right|_{A=0,~\phi=v}\\&&=f^2 v \frac{1}{4\pi} \log
\left(\frac{M^2}{F^2}\right)\delta^2(k)+\mathcal{O}(M^{-2}), 
\end{eqnarray} 
and their contribution to the logarithm of the determinant is
\begin{eqnarray}
&&\int \frac{d^2k}{(2\pi)^2}(\phi(k)-v\delta^2(k))\left.\frac{\delta}
{\delta\phi(k)}\log\left(\det_{reg} [K^\dagger K_{A,\phi}]\right)\notag
\right|_{A=0,\phi=v} + 
\mathrm{h.~c.}\\
&&=\int d^2x \left\{v(\phi(x)-v)+v(\phi^\dagger(x)-v)\right\} 
\frac{f^2}{2\pi}\log\left(\frac{M^2}{F^2}\right),
\end{eqnarray}
with $\delta^2$ the two-dimensional Dirac delta function.
The second derivative reads
\begin{eqnarray*} 
&&\left.\frac{\delta}{\delta\phi^\dagger(q)}\frac{\delta}{\delta\phi(k)} 
\log\left(\det_{reg}[K^\dagger K(\phi,A)]\right)
\right|_{A=0,~\phi=v}\label{ddregM} \\ 
&&=\frac{f^2}{2\pi}\delta(k-q) 
\log\left(\frac{M^2}{F^2}\right)+\mathcal{O}(M^{-2}), 
\end{eqnarray*} 
which gives the following contribution to the logarithm of the determinant: 
$$
\frac{f^2}{2\pi}\int 
d^2x|\phi(x)-v|^2\log\left(\frac{M^2}{F^2}\right).
$$
The contribution of the first and second order derivatives can
be added to give the second term in (\ref{diagPVi}),
which represent a renormalization of the Higgs mass.
\subsection{Functional derivatives with respect to the vector field} 
The first derivative with respect to the vector field 
$$
\left.\frac{\delta}{\delta A_\sigma (k)} \log\left(\det_{reg}[K^\dagger 
K_{A,\phi}]\right)\right|_{A=0,~\phi=v}
$$
does not vanish and gives a contribution to the determinant of the form: 
\begin{equation}
\frac{e}{2}\log\left(\frac{M^2}{F^2}\right)\int 
\frac{d^2x}{2\pi}\epsilon_{\mu\sigma}\partial_\mu 
A_\sigma(x).\label{diagepsdA}
\end{equation} 
Note that $\int \frac{d^2x}{2\pi}\epsilon_{\mu\sigma}\partial_\mu 
A_\sigma(x)$ is just the topological charge. For 
small perturbations around the vacuum with usual boundary conditions 
(like infinite space and finite energy, see \cite{R}), this integral 
is equal to zero. However this is not true in general and 
in the present case this integral is equal to $-1$. 

\subsection{Photon mass term with Pauli-Villars regularization} 
The regularization 
procedure we used is gauge invariant. As this is not 
completely trivial, we shall now check it evaluating the one-loop corrections 
to the photon propagator. This can be calculated with the second 
derivative of (\ref{regM}) with respect to $A$ or by evaluating the 
corresponding Feynman diagrams. Surprisingly, the result, in the limit  
where the photon momentum $q$ goes to zero reads: 
\begin{eqnarray} 
\left.\lim_{q^2\to 0}\frac{\delta}{\delta A_\sigma 
(k)}\frac{\delta}{\delta A_\rho (q)}\log\left(\det_{reg}[K^\dagger 
K_{A,\phi}]\right)\right|_{A=0,~\phi=v}&=& \frac{e^2}{4\pi}\delta_{\mu 
\nu}\label{phot}, 
\end{eqnarray} 
That is we get a mass term for the photon. 
It is important to note that in chiral gauge theories regularized with  
a non-gauge invariant procedure, this is a common feature.
But here, unlike for instance in dimensional 
regularization, the regulator term $M^2$ is gauge 
invariant, and such a problem should not arise. Indeed 
performing further calculations, we can see that every term of the 
scalar field covariant derivative $|(\partial_\mu -ieA_\mu)\phi|^2$ 
receives a finite contribution from the fermion loop, so that this vector 
field mass term can be absorbed in a gauge invariant expression. 
This confirms that Pauli-Villars regularization preserves chiral 
gauge invariance.

In the remaining of the section, the calculation of the fermionic 
contribution to photon propagator is 
presented in more detail. Three diagrams are divergent or 
constant when the photon momentum goes to zero. We do not present the 
full calculation by second derivatives of the action but only 
these three main contributions. 
The first diagram is the 1-vertex loop with $\frac{e^2}{4}A_\mu^2$ 
interaction: 
\begin{eqnarray} 
\psset{unit=0.5cm} 
\begin{pspicture}(0,0)(5,2) 
\rput(2.4,1.2){$\vec{p}$} 
\psline[arrowsize=0.2]{->}(2.5,1.8)(2.7,1.8) 
\rput(3.5,0.8){$\frac{e^2}{4}$} 
\rput(1,0.7){$\vec{q}$} 
\pnode(0,0.5){M_1} 
\rput(2.4,0){\psCoil[coilaspect=0,coilheight=2,coilwidth=0.3, linewidth=1pt]{90}{1500}} 
\rput(2.6,0){\psCoil[coilaspect=0,coilheight=2,coilwidth=0.3, linewidth=1pt]{-1500}{-90}} 
\pscurve[linearc=0.05,linewidth=1pt](2.5,0)(1.9,1.3)(2.5,1.8)(3.1,1.3)(2.5,0) 
\cnode[fillstyle=solid,fillcolor=black](2.5,0){.1}{M_2} 
\end{pspicture} 
&=&\frac{e^2}{4}\int\frac{d^2p}{(2\pi)^2}\tr(\identity) 
\delta_{\mu\nu}\left(\frac{1}{p^2+F^2}-\frac{1}{p^2+F^2+M^2}\right)\notag\\ 
&=&\frac{e^2}{4} \frac{\delta_{\mu\nu}}{2\pi} 
\left[\log\left(\frac{M^2}{F^2}\right)+\mathcal{O}(M^{-2})\right]. 
\end{eqnarray} 
The second diagram is the 2-vertexes loop with $-ie\gamma_5 
A_\mu\partial_\mu$ interaction: 
\begin{eqnarray} 
\psset{unit=0.5cm} 
\begin{pspicture}(0,0)(5,1) 
\rput(2.5,0.4){$\vec{p}$} 
\psline[arrowsize=0.2]{->}(2.5,1)(2.6,1) 
\rput(0.6,-0.6){$\vec{q}$} 
\cnode[fillstyle=solid,fillcolor=black](1.5,0){.1}{M_2} 
\cnode[fillstyle=solid,fillcolor=black](3.5,0){.1}{M_2} 
\pscircle[linewidth=1pt](2.5,0){1} 
\rput(3.4,0){\psCoil[coilaspect=0,coilheight=2,coilwidth=0.3, linewidth=1pt]{90}{990}} 
\rput(1.6,0){\psCoil[coilaspect=0,coilheight=2,coilwidth=0.3, linewidth=1pt]{-990}{-90}} 
\end{pspicture} 
&=&-\left.\frac{e^2}{2!}\int\frac{d^2p}{(2\pi)^2} 
\left(\frac{\tr((\gamma_5)^2)~ip_\mu}{p^2+F^2+M^2} 
\frac{i(p+q)_\nu}{(p+q)^2+F^2+M^2}\right)\right|_M^{M=0}\notag\\ 
&=&-\frac{e^2}{4} \frac{\delta_{\mu\nu}}{2\pi} 
\left[\log\left(\frac{M^2}{F^2}\right)+\mathcal{O}(M^{-2})+ 
\mathcal{O}(q^{2})\right]. 
\end{eqnarray} 
The integration over $p$ is done by standard techniques. These 
 first two diagrams cancel each other to 
$\mathcal{O}(M^{-2})+\mathcal{O}(q^{2})$; but the third one gives 
some constant contribution. Let us consider the 2-vertexes loop with 
$-ief\phi\gamma_\mu A_\mu$ interaction; $\phi$ is considered to be in 
vacuum configuration $\phi=v$ and $fv=F$: 
\begin{eqnarray} 
\psset{unit=0.5cm} 
\begin{pspicture}(0,0)(5,1) 
\rput(2.5,0.4){$\vec{p}$} 
\psline[arrowsize=0.2]{->}(2.5,1)(2.6,1) 
\psline[linewidth=1pt,linestyle=dashed](0.5,1)(1.5,0) 
\psline[linewidth=1pt,linestyle=dashed](4.5,1)(3.5,0) 
\cnode[fillstyle=solid,fillcolor=black](0.5,1){.1}{M_2} 
\cnode[fillstyle=solid,fillcolor=black](4.5,1){.1}{M_2} 
\rput(0.6,-0.6){$\vec{q}$} 
\cnode[fillstyle=solid,fillcolor=black](1.5,0){.1}{M_2} 
\cnode[fillstyle=solid,fillcolor=black](3.5,0){.1}{M_2} 
\pscircle[linewidth=1pt](2.5,0){1} 
\rput(3.4,0){\psCoil[coilaspect=0,coilheight=2,coilwidth=0.3, linewidth=1pt]{90}{990}} 
\rput(1.6,0){\psCoil[coilaspect=0,coilheight=2,coilwidth=0.3, linewidth=1pt]{-990}{-90}} 
\end{pspicture} 
&=&-\left.\frac{e^2}{2!}\int\frac{d^2p}{(2\pi)^2} 
\left(\frac{\tr(\gamma_\mu\gamma_\nu)}{p^2+F^2+M^2} 
\frac{F^2}{(p+q)^2+F^2+M^2}\right)\right|_M^{M=0}\notag \\ 
&=&\frac{e^2}{2}\frac{\delta_{\mu\nu}}{2\pi} 
\left[1+\mathcal{O}(M^{-2})+ \mathcal{O}(q^{2})\right]. 
\end{eqnarray} 
This gives equation (\ref{phot}),
which violates at first sight the chiral gauge invariance. 
Let us consider now other terms involving scalar and vector fields 
get such contributions, namely $|\partial_\mu \phi|^2$, 
\mbox{$-ieA_\mu(\phi^*\partial_\mu \phi-\phi \partial_\mu \phi^*)$}. Let us 
consider the diagram with two vertexes 
$f\gamma_\mu\partial_\mu(\Re(\phi)+i\gamma_5\Im(\phi))$:
\begin{eqnarray} 
\psset{unit=0.5cm} 
\begin{pspicture}(0,0)(5,1) 
\rput(2.5,0.4){$\vec{p}$} 
\psline[arrowsize=0.2]{->}(2.5,1)(2.6,1) 
\psline[linewidth=1pt,linestyle=dashed](0,0)(1.5,0) 
\psline[linewidth=1pt,linestyle=dashed](5,0)(3.5,0) 
\rput(0.6,-0.6){$\vec{q}$} 
\cnode[fillstyle=solid,fillcolor=black](1.5,0){.1}{M_2} 
\cnode[fillstyle=solid,fillcolor=black](3.5,0){.1}{M_2} 
\pscircle[linewidth=1pt](2.5,0){1} 
\end{pspicture} 
&=&\left.\frac{f^2}{2!}\int\frac{d^2p}{(2\pi)^2} 
\left(\frac{\tr(\gamma_\mu\gamma_\nu)}{p^2+F^2+M^2}\frac{q_\mu 
q_\nu}{(p+q)^2+F^2+M^2}\right)\right|_M^{M=0}\notag \\ 
&=&\frac{\delta_{\mu\nu}q^2}{4\pi v^2}+ 
\mathcal{O}(M^{-2})+\mathcal{O}(q^{4}),\label{diagcov1} 
\end{eqnarray} 
which gives a contribution  
\begin{equation} 
\frac{1}{4\pi}\frac{|\partial_\mu \phi|^2}{v^2}\label{dphi2} 
\end{equation} 
 to the effective action. The next diagram is the mixed one and 
contains one vertex 
$f\gamma_\mu\partial_\mu(\Re(\phi)+i\gamma_5\Im(\phi))$ and one 
$-ief\phi\gamma_\mu A_\mu$; the product of these vertexes gives two 
terms:  $$2ief^2A_\mu (\phi^*\partial_\mu\phi 
-\phi\partial_\mu\phi^*)+ef^2\varepsilon_{\mu\nu}(\phi^*\partial_\mu\phi 
+\phi\partial_\mu\phi^*)A_\nu.$$  We drop the second one, which is not 
part of the scalar covariant derivative, and which is gauge invariant 
(up to total derivative): 
 
\begin{eqnarray} 
\psset{unit=0.5cm} 
\begin{pspicture}(0,0)(5,1) 
\rput(2.5,0.4){$\vec{p}$} 
\psline[arrowsize=0.2]{->}(2.5,1)(2.6,1) 
\psline[linewidth=1pt,linestyle=dashed](0,0)(1.5,0) 
\psline[linewidth=1pt,linestyle=dashed](4.5,1)(3.5,0) 
\rput(0.6,-0.6){$\vec{q}$} 
\cnode[fillstyle=solid,fillcolor=black](1.5,0){.1}{M_2} 
\cnode[fillstyle=solid,fillcolor=black](3.5,0){.1}{M_2} 
\pscircle[linewidth=1pt](2.5,0){1} 
\rput(3.4,0){\psCoil[coilaspect=0,coilheight=2,coilwidth=0.3, linewidth=1pt]{90}{990}} 
\end{pspicture} 
&=&\left.\frac{1}{2!}\int\frac{d^2p}{(2\pi)^2} 
\left(\frac{2ief^2}{p^2+F^2+M^2} 
\frac{i(\delta_{\mu\nu}q_\nu+\delta_{\mu\nu}q'_\nu)}{(p+q)^2+F^2+M^2}\right) 
\right|_M^{M=0}\notag 
\\ &=&\frac{ie}{2\pi} 
\frac{i(\delta_{\mu\nu}q_\nu+\delta_{\mu\nu}q'_\nu)}{2v^2} 
\left[1+\mathcal{O}(M^{-2})+\mathcal{O}(q^{2})\right],\label{diagcov2} 
\end{eqnarray} 
with $q$ the momentum of incoming scalar field $\phi$ and $q'$ the  
out-coming one.  
This gives a contribution to the scalar-gauge effective action of 
the form: 
\begin{equation} 
\frac{ie}{2\pi v^2}A_\mu(\phi \partial_\mu\phi^*-\phi^* 
\partial_\mu\phi).\label{dphi3} 
\end{equation} 
It is now possible to resume the terms (\ref{phot}, \ref{dphi2}, 
\ref{dphi3}) in a manifestly gauge invariant term $\frac{1}{4\pi 
v^2}|(\partial_\mu-ieA_\mu)\phi|^2$ to be added to the initial scalar 
covariant derivative $\frac{1}{2}|(\partial_\mu-ieA_\mu)\phi|^2$ and 
the photon acquires a mass
\begin{equation}
m_W^{1-loop}=\sqrt{e^2v^2+\frac{e^2}{2\pi}}.\label{mw}
\end{equation}
This mass can be expressed with the dimensionless parameters 
(\ref{dimlessparam}) as 
$\frac{m_w^{1-loop}}{m_H}=\sqrt{\frac{1}{2\mu^2}+\frac{1}{4\pi^2}}$, 
which does not depend on the fermion mass. Note that in the case of 
massless fermions, a similar phenomenon appears (Schwinger mechanism 
\cite{2dqed}). 
\subsection{Determinants of small fluctuations} 
We are checking here if the counterterms mentioned before are 
sufficient to get a finite determinant. In order to be able to do it 
analytically we will only consider some small constant perturbation 
and calculate the ratio of the determinants in (\ref{regM}): First 
let us take $\phi=v+\delta\phi$, $A=0$, and note that $\delta 
F=f\delta\phi$: 
\begin{eqnarray*} 
 &&\log\left(\frac{\det[K^\dagger K(\phi,A)]}{\det[K^\dagger 
K_{vac}]}\frac{\det[K^\dagger K_{vac} + M^2]}{\det[K^\dagger 
K(\phi,A)+M^2]}\right)\\ 
&&=\log\left(\frac{\det[\identity(-\partial_0^2 - \partial_1^2 + 
(F+\delta F)^2)]}{\det[\identity(-\partial_0^2 - \partial_1^2 + 
F^2)]} \frac{\det[\identity(-\partial_0^2 - \partial_1^2 + 
F^2+M^2)]}{\det[\identity(-\partial_0^2 - \partial_1^2 + (F+\delta 
F)^2+M^2)]}\right). 
\end{eqnarray*} 
In momentum space, we can rewrite the last expression as
$$ 
\left[\int \frac{d^2k}{(2\pi^2)}\log\left(\frac{(k^2+(F+\delta 
F)^2)^2}{(k^2+F^2)^2}\frac{(k^2+F^2+M^2)^2}{(k^2+(F+\delta 
F)^2+M^2)^2}\right)\right],$$ 
which can be easily calculated to give 
$$\frac{f^2}{2\pi}((v+\delta\phi)^2-v^2) 
\log\left(\frac{M^2}{F^2}\right).$$ 
As the logarithm of the determinant is the sum of all one loop 
diagrams, we have to make subtractions at this level. Clearly the second 
term of the counterterm (\ref{diagPVi}) removes the 
divergence of this determinant. Then if we take $\phi=v$, 
$A_\mu=\delta A_\mu$ small constant perturbations; it is easy to 
perform the same calculations to see that no divergent term occurs. 
Similarly if we take simultaneously $\phi=v+\delta\phi$ and 
$A_\mu=\delta A_\mu$ the calculation is more complicated but we 
recover once again the previous divergences. However we can see that, 
taking a specific configuration where 
$\varepsilon_{\mu\nu}\partial_\mu A_\nu$ is constant, $\phi=v$ and 
$A=0$, we find a divergent contribution of the form: 
$$\frac{e}{4\pi}\varepsilon_{\mu\nu}\partial_\mu 
A_\nu\log\left(\frac{M^2}{F^2}\right),$$ 
which is subtracted exactly by the counterterm (\ref{diagepsdA}). 
\subsection{Equivalence between Pauli-Villars and partial wave \\counterterms} 
For small constant background fields we may compare the partial wave 
counterterm (\ref{diagpv}) and the Pauli-Villars one (\ref{diagPVi})
The difference $\delta S^{UV}_{count}$ between them  
relates the different cutoffs $M$ and $\frac{2L}{R}$:
$$\delta S^{UV}_{count}(M,\frac{2L}{R})=\int_0^R\left(f^2\left(|\phi(r)|^2-
v^2\right)+\frac{e}{2}\varepsilon_{\mu,\nu} 
\partial_\mu A_\nu(r)\right)d^2r\frac{1}{2\pi} 
\left[1+log\left(\frac{4L^2}{M^2R^2}\right)\right].$$ 
For any background that approach vacuum at infinity, it can
be shown that the Pauli-Villars counterterms are equivalent to the
partial wave ones.
We introduce a Pauli-Villars regulator in the partial wave counterterm  
(\ref{diagpv}): 
 \begin{eqnarray*} 
&&\sum_{m=-L}^L\int_0^R 2\pi r\tr[G_F^m(r,r)] 
h(r)dr\\ 
&&=\lim_{M\to\infty}\sum_{m=-L}^L\int_0^R 2\pi r~ 
\left(\tr[G_F^m(r,r)]-\tr[G_M^m(r,r)]\right) 
h(r)dr 
\end{eqnarray*} 
with $h(r)=\left(f^2\phi^2+\frac{e}{2}\varepsilon_{\mu\nu} 
\partial_\mu A_\nu\right)$ and $G_F$ the Green's function for a particle of 
mass $F$ given in equation (\ref{Gm}). 
The sum over $m$ is now convergent and we can take $L\to\infty$. The sum 
of the Green's functions reads: 
$$\sum_{m=-\infty}^\infty 
\left(\tr[G_F^m(r,r)]-\tr[G_M^m(r,r)]\right)=\frac{1}{\pi}\sum_{m=-\infty}^
\infty \left(I_m(Fr)K_m(Fr)-I_m(Mr)K_m(Mr)\right). 
$$
Note that the second term in the Green's function (\ref{Gm}) can be dropped if 
the potential decreases fast enough at infinity, which is the case here. 
 
We use the following sum rule for Bessel functions: $\sum_{m=-\infty}^\infty 
I_m(Fr)K_m(Fr')=K_0(F(r-r'))$, therefore, we rewrite the previous expression with 
two different radii: 
\begin{eqnarray*} 
&=&\frac{1}{\pi}\lim_{r'\to r}\sum_{m=-\infty}^\infty \left(I_m(Fr)K_m(Fr')
-I_m(Mr)K_m(Mr')\right)\\ 
&=&\frac{1}{\pi}\lim_{r'\to r}[K_0(F(r-r'))-K_0(M(r-r'))]. 
\end{eqnarray*} 
For small r, we have $K_0(r)\sim -\ln(r)$ and $[K_0(F(r-r'))-K_0(M(r-r'))]=
ln\left[\frac{M}{F}\right]$. The limit $r'\to r$ is trivial and we get for 
the whole counterterm: 
\begin{equation} 
\frac{1}{2\pi}\ln\left[\frac{M^2}{F^2}\right]\int_0^Rh(r)rdr, 
\end{equation} 
which is precisely the counterterm in the Pauli-Villars scheme (\ref{diagPVi}). 
\section{One-loop divergences in partial waves} 
The divergent diagrams studied in the framework of Pauli-Villars 
regularization, see Appendix A.2, A.3, A.4, can be 
recalculated with partial waves for a constant background.
 Their sum is expressed in equation (\ref{diagpv}) and we
perform the integration in the following. We have:
\begin{equation} 
G^m(r,r)=\frac{\identity}{2\pi}\left(I_m(Fr)K_m(Fr)- 
\frac{K_m(FR)}{I_m(FR)}I_m(Fr)^2 \right)\label{Grr}, 
\end{equation} 
which may be simplified using asymptotic 
expansions in order for Bessel functions. As the divergences are 
coming from large $m$, this approximation takes care of the necessary 
contributions: 
\begin{eqnarray} 
I_m(Fr)&=&\frac{1}{\sqrt{2\pi}}\frac{1}{(m^2+F^2r^2)^{1/4}} 
\exp\left[\sqrt{m^2+F^2r^2}-m~ 
\mathrm{arcsinh}\left(\frac{m}{Fr}\right)\right]\label{BesselasI},\notag\\ 
K_m(Fr)&=&\sqrt{\frac{2}{\pi 
}}\frac{1}{(m^2+F^2r^2)^{1/4}}\exp\left[-\sqrt{m^2+F^2r^2}+m~ 
\mathrm{arcsinh}\left(\frac{m}{Fr}\right)\right]\label{BesselasK}. 
\end{eqnarray} 
The second term in the propagator (\ref{Grr}) is very small if 
$R\gg1$ and can be neglected. If the background is supposed to be constant,  
it can be taken out of the integral, (\ref{diagpv}) becomes
\begin{eqnarray} 
&&\left(f^2\left(|\phi|^2-v^2\right)+\frac{e}{2}\varepsilon_{\mu,\nu}\partial_\mu 
A_\nu\right)\sum_{m=-L}^L\int_0^R 2\pi r \notag
\frac{1}{2\pi\sqrt{m^2+F^2r^2}}dr\\ &=& 
\left(f^2\left(|\phi|^2-v^2\right)+\frac{e}{2}\varepsilon_{\mu,\nu}\partial_\mu 
A_\nu\right)\sum_{m=-L}^L 
\frac{1}{F^2}\left(\sqrt{m^2+F^2r^2}-m\right), 
\end{eqnarray}  
where the sum can be converted to an integral: 
\begin{eqnarray} 
&& \left(f^2\left(|\phi|^2-v^2\right)+\frac{e}{2}\varepsilon_{\mu,\nu}
\partial_\mu 
A_\nu\right)2 \int_{0}^L \notag
\frac{1}{F^2}\left(\sqrt{m^2+F^2r^2}-m\right)dm\\ 
&\simeq&\left(f^2\left(|\phi|^2-v^2\right)+\frac{e}{2}\varepsilon_{\mu,\nu}
\partial_\mu 
A_\nu\right)\frac{R^2}{2}\left[1+\log\left(\frac{4L^2}{F^2R^2} 
\right)\right]. 
\end{eqnarray}  
rewriting 
$\left(f^2\left(|\phi|^2-v^2\right)+\frac{e}{2}\varepsilon_{\mu,\nu}
\partial_\mu A_\nu\right)$ as an integral over space lead to (\ref{divpv}).
\subsection{Photon mass term in partial wave}
Finally we recalculate the photon propagator in partial waves. 
The fermionic contribution to the photon propagator comes from three diagrams. 
The first one reads 
\begin{equation} 
\psset{unit=0.7cm} 
\begin{pspicture}(0,0)(5,2) 
\psline[arrowsize=0.2]{->}(2.5,1.8)(2.7,1.8) 
\rput(3.4,0.6){$\frac{e^2}{4}$} 
\rput(2.5,-0.5){$\vec{r}$} 
\pnode(0,0.5){M_1} 
\rput(2.4,0){\psCoil[coilaspect=0,coilheight=2,coilwidth=0.3, linewidth=1pt]{90}{1500}} 
\rput(2.6,0){\psCoil[coilaspect=0,coilheight=2,coilwidth=0.3, linewidth=1pt]{-1500}{-90}} 
\pscurve[linearc=0.05,linewidth=1pt](2.5,0)(1.9,1.3)(2.5,1.8)(3.1,1.3)(2.5,0) 
\cnode[fillstyle=solid,fillcolor=black](2.5,0){.1}{M_2} 
\end{pspicture} 
=\frac{e^2}{4}\sum_m\int_0^R 2\pi r dr G^m(r,r) \tr(\identity)A_\mu^2(r). 
\end{equation} 
This integration is precisely the same as (\ref{diagpv}), and the 
result is: 
\begin{equation} 
\frac{e^2}{4}\int_0^R A^2_\mu 
d^2r~\frac{1}{2\pi}\left[1+\log\left(\frac{4L^2}{F^2R^2}\right)\right]. 
\end{equation} 
The second diagram is ~~ 
\psset{unit=0.7cm} 
\begin{pspicture}(0,0)(5,1) 
\rput(1.2,0.6){$\vec{r}$} 
\psline[arrowsize=0.2]{->}(2.5,1)(2.6,1) 
\rput(3.8,0.6){$\vec{r'}$} 
\cnode[fillstyle=solid,fillcolor=black](1.5,0){.1}{M_2} 
\cnode[fillstyle=solid,fillcolor=black](3.5,0){.1}{M_2} 
\pscircle[linewidth=1pt](2.5,0){1} 
\rput(3.4,0){\psCoil[coilaspect=0,coilheight=2,coilwidth=0.3, linewidth=1pt]{90}{990}} 
\rput(1.6,0){\psCoil[coilaspect=0,coilheight=2,coilwidth=0.3, linewidth=1pt]{-990}{-90}} 
\end{pspicture} 
~~with vertices $ie\gamma_5 A_\mu\partial_\mu$. 
\vspace{0.7cm} 
In our case, 
$A_\mu\partial_\mu=A_r\partial_r+A_\theta\frac{1}{r}\partial_\theta$ 
with $A_r=0$ and $\frac{1}{r}\partial_\theta$ is replaced by 
$\frac{m}{r}$ for the partial wave $m$. We further assume that 
$A_\theta$ is constant over all space. The above diagram gives 
\begin{equation} 
-\frac{1}{2!}e^2 \tr(\gamma_5^2)\sum_m\int d^2r 
d^2r'A_\theta\frac{m}{r}G^m(r,r')A_\theta\frac{m}{r'}G^m(r',r),\label{108} 
\end{equation} 
where $G^m(r,r')$ is given by (\ref{Grr}).
We are interested in large $m$ contributions, and therefore we use 
the asymptotic formulas (\ref{BesselasI}, \ref{BesselasK}) for Bessel 
functions in the propagator. We are also interested in the limit 
$R\gg 1$, therefore we drop once again the second term in the 
propagator. After some calculations we get : 
\begin{eqnarray} 
G^2(r,r')=\frac{1}{4(2\pi)^2}\frac{1}{\sqrt{m^2+F^2r^2}} 
\frac{1}{\sqrt{m^2+F^2r'^2}} 
\left\{\exp\left[2g(m,r,r')\right]\theta(r'-r)\right.\notag\\ 
\left.+~\exp[2 g(m,r',r)]\theta(r-r')\right\}.\label{Grr2as} 
\end{eqnarray} 
with 
$$g(m,r,r')=\left(\sqrt{m^2+F^2r^2}-\sqrt{m^2+F^2r'^2}- 
m\,\mathrm{arcsinh}\frac{m}{Fr}+m\,\mathrm{arcsinh}\frac{m}{Fr'}\right).$$ 
The dominant contribution comes from diagrams with $r\cong r'$. 
Expanding in powers of $r-r'$ and performing the integrations, we get for 
(\ref{108}):
\begin{eqnarray} 
&&-\frac{e^2}{4}A_\theta^2\sum_m \int_0^R\frac{m^2 \notag
dr}{m^2+F^2r^2}\left\{\int_r^R 
dr'\exp\left[2\frac{\sqrt{m^2+F^2r^2}(r-r')}{r}\right]\right.\\ 
&&\left.+ \int_0^r \notag
dr'\exp\left[2\frac{\sqrt{m^2+F^2r^2}(r'-r)}{r}\right]\right\} \\
&=&-\frac{e^2}{4}A_\theta^2\sum^L_{m=-L}m^2\left[\frac{1}{F^2m}- 
\frac{1}{F^2\sqrt{m^2+F^2R^2}}\right]\notag\\ 
&\simeq&-\frac{e^2}{4}A_\theta^2~2\int^L_0m^2\left[\frac{1}{F^2m}- 
\frac{1}{F^2\sqrt{m^2+F^2R^2}}\right]dm\notag\\ 
&=&-\frac{e^2}{2}A_\theta^2\left[\frac{1}{4}R^2 
\left(-1+\log\left(\frac{4L^2}{F^2R^2}\right)\right)+ 
\mathcal{O}(L^{-2})\right]\notag\\ &\simeq&-\frac{e^2}{4}\int_0^R A_\mu^2 
d^2r~\frac{1}{2\pi}\left[-1+\log\left(\frac{4L^2}{F^2R^2}\right)\right]. 
\end{eqnarray} 
The third diagram is:~~ 
\begin{pspicture}(0,0)(5,1) 
\rput(3.7,-0.5){$\vec{r'}$} 
\psline[arrowsize=0.2]{->}(2.5,1)(2.6,1) 
\psline[linewidth=1pt,linestyle=dashed](0.5,1)(1.5,0) 
\psline[linewidth=1pt,linestyle=dashed](4.5,1)(3.5,0) 
\cnode[fillstyle=solid,fillcolor=black](0.5,1){.1}{M_2} 
\cnode[fillstyle=solid,fillcolor=black](4.5,1){.1}{M_2} 
\rput(1.3,-0.5){$\vec{r}$} 
\cnode[fillstyle=solid,fillcolor=black](1.5,0){.1}{M_2} 
\cnode[fillstyle=solid,fillcolor=black](3.5,0){.1}{M_2} 
\pscircle[linewidth=1pt](2.5,0){1} 
\rput(3.4,0){\psCoil[coilaspect=0,coilheight=2,coilwidth=0.3, linewidth=1pt]{90}{990}} 
\rput(1.6,0){\psCoil[coilaspect=0,coilheight=2,coilwidth=0.3, linewidth=1pt]{-990}{-90}} 
\end{pspicture} 
~~with vertices $-ief\phi\gamma_\mu A\mu$. 
\vspace{0.5cm} 
\begin{eqnarray} 
-\frac{1}{2!}\sum_m \frac{e^2F^2}{v^2} 
\tr(\gamma_\nu^2)A_\mu^2\phi^2\int r dr r'dr'(2\pi)^2 (G^m(r,r'))^2. 
\end{eqnarray} 
Using an asymptotic expression for the propagator (\ref{Grr2as}) as 
before, and doing the integration in a similar way, we get: 
\begin{eqnarray} 
\frac{e^2}{4v^2\pi}\int_0^R \phi^2A_\mu^2 d^2r.
\end{eqnarray} 
The very same way we can recalculate the diagrams (\ref{diagcov1}, 
\ref{diagcov2}) to find respectively  $\frac{1}{4\pi 
v^2}\int_0^R|\partial_\mu \phi|^2 d^2r$ and $\frac{-ie}{4\pi 
v^2}\int_0^R A_\mu (\phi^*\partial_\mu\phi-\phi\partial_\mu\phi^*) 
d^2r$. These three last expressions can be rewritten into a covariant 
derivative $\frac{1}{4\pi v^2}|(\partial_\mu-ieA_\mu)\phi|^2$. 
The first two do not cancel completely and a term 
$\frac{e^2}{4\pi}A_\mu^2$ needs to be subtracted from the action, to get a 
gauge invariant regularization (\ref{Scontinfra}). After this the physical
vector boson mass is given by (\ref{mw}).
\section{Exchanging the limits} 
Two limits were considered in the determinant calculation, the limit of 
infinite volume ($R\to\infty$) and the limit of infinite cutoff in the 
sum over the partial waves ($L\to\infty$).
The order of limits specified in equation (\ref{detrenpw}), that is to say
take $L\to\infty$ first and then $R\to \infty$, is essential. 
In this Appendix, we calculate the determinant in 
the case of vanishing instanton 
core size\footnote{Taking a zero instanton core size lead 
to normalization problem for the zero-mode. This is not essential for our 
purposes, and it is possible to reproduce all these calculations more 
rigorously considering a ``step'' core, where $f(r)=A(r)=0$, 
$r<\delta$; and then consider $\delta\to0$. However the calculations 
are tedious and the same conclusions remain.} 
and consider what would happen we commute the limits. 
In this simple case 
everything can be done analytically; the counterterm  to the 
scalar field mass vanishes, because of zero core size. The result 
for the sum over non-zero partial wave $m$ should be finite after 
removing counterterms related to vector fields.  
The solutions in the case $n=-1$ with 
boundary conditions (\ref{cb}) are: 
\begin{eqnarray} 
\Psi_L^{m,inst}(r)=I_{m-1/2}(Fr)\frac{\Gamma(m+1/2)}{\Gamma(m+1)}, 
&&\quad\Psi_L^{m,vac}(r)=I_{m}(Fr),\notag \\ 
\Psi_R^{m,inst}(r)=I_{m+1/2}(Fr)\frac{\Gamma(m+3/2)}{\Gamma(m+1)}, 
&&\quad \Psi_R^{m,vac}(r)=I_{m}(Fr). 
\end{eqnarray} 
Using 
$$I_m(r) \overset{r\to \infty} \to \frac{e^r}{\sqrt{2\pi 
r}}\left(1-\frac{4m^2-1}{8r}+\mathcal{O}(r^{-2})\right),$$ 
the determinant for $R\to \infty$ is given by: 
\begin{eqnarray*} 
\det[M^m]&=&\frac{\Psi_L^{m,inst}(\infty)\Psi_R^{m,inst} 
(\infty)}{\Psi_L^{m,vac}(\infty)\Psi_R^{m,vac}(\infty)} 
=\frac{\Gamma(m+1/2)\Gamma(m+3/2)}{\Gamma(m+1)^2} 
\left(1+\mathcal{O}(r^{-1})\right)\\ 
&\overset{m\to\infty}=&1+\frac{1}{4m}+\mathcal{O}(m^{-2}). 
\end{eqnarray*} 
Clearly $\prod_m \left(1+\frac{1}{4m}\right)$ diverges. Note that 
using this method for the complete numerical calculation, the very 
same divergence remains after removing the ultraviolet counterterms. 
 
With the second method, using asymptotic expansion (\ref{BesselasK}) for large $m$ 
and finite radius, we get: 
\begin{eqnarray*} 
\det[M^m]&\overset{m\gg 
1}\to&\frac{\Gamma(m+1/2)\Gamma(m+3/2)}{\Gamma(m+1)^2}\left(1-\frac{1}{4m}+ 
\mathcal{O}(m^{-2})+\mathcal{O}(r^2m^{-3})\right)\\&=&1+\mathcal{O}(m^{-2}). 
\end{eqnarray*} 
which gives a convergent product. This shows that, also in this 
simple case, we have to perform the sum over $m$ to infinity 
before taking $R\to\infty$, otherwise we do not get a sensible answer.  
\section{Determinant at small fermion mass} 
Observation of numerical results shows a power law behavior of the 
determinant for small fermion mass. More precisely this power law 
comes from the partial determinant $\sqrt{\det'M^0_+}$, where we 
remove the zero-mode. It is also this contribution that provides the 
dimension of $\textrm{mass}^{-1}$ for the determinant. It would be interesting 
to find this behavior by analytical calculations. To this end we will use another method \cite{coleman} than (\ref{modezero}) to remove the 
 zero eigenvalue. 

The zero mode wave function which vanishes at the 
boundary is noted $\Psi_0(r)$ and $\Phi_0(r)$ shall be
the other solution of the second order 
differential equation (\ref{equamodezero}): 
\begin{eqnarray} 
\Psi_0(r)&=&e^{-\int_0^rdr'g(r')},\quad \mathrm{with}\quad 
g(r)=Ff(r)+\frac{e}{2}A(r),\label{psi0}\\ 
\Phi_0(r)&=&e^{-\int_0^rdr'g(r')}\int_a^r 
\frac{dr'}{r'}e^{\int_0^{r'}dr''2 g(r'')}\label{phi0}. 
\end{eqnarray} 
This last solution is not normalizable and the constant $a$ which 
defines the integral is arbitrary. We consider the system to be in a 
spherical box of radius $R$. The actual solution, which vanishes 
at the boundary, is not $\Psi_0(r)$ anymore but $\Psi_\lambda(r)$, 
which has a non-zero eigenvalue.  $\Psi_\lambda(r)$ can be found 
with the help of perturbation theory:
\begin{equation} 
\Psi_\lambda(r)=\Psi_0(r)-\lambda \int_0^r 
t\,dt\left[\Phi_0(r)\Psi_0(t)-\Psi_0(r)\Phi_0(t)\right]\Psi_0(t), 
\end{equation} 
where the two solutions (\ref{psi0}, 
\ref{phi0}) are normalized so that their Wronskian is $1/r$ exactly. We replace 
$\Psi_\lambda(R)=0$ in the previous equation, this yields 
$$\lambda = \frac{h(R)}{\Psi_0(R)},\quad h(R)=\int_0^R 
t\,dt\left[\Phi_0(R)\Psi_0(t)-\Psi_0(R)\Phi_0(t)\right]\Psi_0(t).$$ 
Then the determinant with lowest eigenvalue omitted is 
$$\det'(M^0_+)=\frac{\Psi_0(R)}{\Psi_{vac}(R)}\frac{1}{\lambda} 
=\frac{h(R)}{\Psi_{vac}(R)}.$$ 
In order to find an analytical approximation for this last 
expression, we use the following approximate profile for the instanton: 
\begin{eqnarray} 
eA(r)&=&\left\{\begin{array}{cc} e^2r, & r \leq 1/e,\\1/r, & r 
> 1/e,\end{array}\right.\notag\\ Ff(r)&=&\left\{\begin{array}{cc} 
eFr, & r \leq 1/e,\\ F, & r > 1/e.\end{array}\right. \label{pro}
\end{eqnarray} 
Note that the powers of $e$ are introduced for dimensional reasons, 
the asymptotic behavior is exact and the behavior near the center 
is closely resembling the instanton core. The solutions 
(\ref{psi0}, \ref{phi0}) become 
\begin{eqnarray*} 
\Psi_0(r)&=&\left\{\begin{array}{cc} 
\exp\left(-\frac{1}{4}e(e+2F)r^2\right), & r \leq 1/e,\\ 
\frac{1}{\sqrt{er}}\exp\left(-\frac{1}{4}+\frac{F}{2e}-Fr\right), & r 
> 1/e,\end{array}\right.\\ 
\Phi_0(r)&=&\left\{\begin{array}{cc}\frac{1}{2}\exp\left( 
-\frac{1}{4}e(e+2F)r^2\right)\qquad \qquad\qquad \qquad\\ 
\times\left[\mathrm{Ei}\left(-\frac{1}{2}e(e+2F)r^2\right)- 
\mathrm{Ei}\left(-\frac{e+2F}{2e}\right)\right] , & r \leq 
1/e,\\  
\sqrt{\frac{e}{r}}\frac{\exp(-Fr+\frac{1}{4}-\frac{F}{2e})}{2F} 
\left(\exp(2Fr)-\exp(2F/e)\right), & r > 
1/e.\end{array}\right.\notag 
\end{eqnarray*} 
Using asymptotic expansions and neglecting parts decreasing as 
$\exp(-Fr)$, the primed determinant yields: 
\begin{eqnarray} 
\det'(M^0_+)&=&\sqrt{\frac{\pi 
e}{2F}}\exp\left(\frac{1}{4}-\frac{F}{2e}\right)\int_0^R 
\Psi_0^2(t)t\,dt \notag\\&= 
&\sqrt{\frac{\pi}{2}}\frac{\exp(1/4)}{2}\frac{1}{e^{1/2}F^{3/2}}+ 
\mathcal{O}(F^{-1/2}).
\end{eqnarray} 
That is to say, for dimensionless variables: 
\begin{equation} 
F\sqrt{\det{M^0_+}} \simeq 0.805 \left(\frac{F}{e}\right)^{1/4}. 
\label{c}
\end{equation} 
It can be compared
to numerical results for the partial determinant $\det{M^0_+}$ with which it 
agrees to few percents. The discrepancy comes from the approximate estimate 
(\ref{pro}) done for the instanton profile. The power law behavior is 
confirmed in figure \ref{f5}. Note that the constant 
$\sqrt{\frac{\pi}{2}}\frac{\exp(1/4)}{2}\simeq 0.805$ in (\ref{c}) 
is not expected to match the
constant found in the fit of figure 5, where the complete determinant was
plotted.

\end{document}